\documentclass[twocolumn,aps,prb,superscriptaddress,longbibliography,floatfix]{revtex4-2}

\usepackage{physics}
\usepackage{times}
\usepackage{graphicx}
\usepackage{amsfonts}
\usepackage{amsmath, amsthm, amssymb, mathrsfs}
\usepackage{empheq}
\usepackage{dsfont}
\usepackage{xcolor}
\usepackage{cancel}
\usepackage{standalone}
\usepackage{stmaryrd}
\usepackage{microtype}
\usepackage{tikz}
\usetikzlibrary{decorations.markings}
\usetikzlibrary{shapes,positioning,arrows}
\usepackage{bm}
\usepackage{siunitx}
\usepackage{slashed}
\usepackage{siunitx}\usepackage[colorlinks=true, allcolors=black, citecolor=blue, linkcolor=blue, urlcolor=blue]{hyperref}
\usepackage[capitalize]{cleveref}
\usepackage{epsfig}

\allowdisplaybreaks

\usepackage{graphicx}
\usepackage[caption=false]{subfig}
\graphicspath{{figures/}}

\DeclareMathOperator{\diag}{diag}
\DeclareMathOperator{\antidiag}{antidiag}
\newcommand{\del}{\partial}

\newcommand{\g}{\gamma}
\newcommand{\gt}{\tilde{\gamma}}

\newcommand{\vv}[1]{{\mathbf{#1}}}

\newcommand{\gs}{\hat{g}}
\newcommand{\gR}{\hat{g}}

\renewcommand{\l}{\left}
\renewcommand{\r}{\right}

\newcommand{\T}{\hat{\mathcal{T}}}
\newcommand{\N}{\hat{\mathcal{N}}}
\newcommand{\B}{\hat{\mathcal{B}}}

\newcommand{\basis}{\vv{e}}

\DeclareMathAlphabet{\mathpzc}{OT1}{pzc}{m}{it}

\newcommand{\ie}{\emph{i.e.}~}
\newcommand{\eg}{\emph{e.g.}~}

\newcommand{\be}{\begin{equation}}
\newcommand{\ee}{\end{equation}}
\newcommand{\bse}{\begin{subequations}}
\newcommand{\ese}{\end{subequations}}
\newcommand{\bal}{\begin{align}}
\newcommand{\eal}{\end{align}}

\newcommand{\curvba}[1]{\hat{\bm{\mathcal{#1}}}}

\newcommand{\tensder}{\Tilde{\mathcal{D}}}

\usepackage[integrals]{wasysym}

\begin{document}

\title{Interface probe for antiferromagnets using geometric curvature}

\author{Tancredi Salamone}
\affiliation{Center for Quantum Spintronics, Department of Physics, NTNU, Norwegian University of Science and Technology, NO-7491 Trondheim, Norway}
\author{Magnus Skjærpe}
\affiliation{Center for Quantum Spintronics, Department of Physics, NTNU, Norwegian University of Science and Technology, NO-7491 Trondheim, Norway}
\author{Henning G. Hugdal}
\affiliation{Center for Quantum Spintronics, Department of Physics, NTNU, Norwegian University of Science and Technology, NO-7491 Trondheim, Norway}
\author{Morten Amundsen}
\affiliation{Nordita, KTH Royal Institute of Technology and Stockholm University, Hannes Alfvéns väg 12, SE-106 91 Stockholm, Sweden}
\affiliation{Center for Quantum Spintronics, Department of Physics, NTNU, Norwegian University of Science and Technology, NO-7491 Trondheim, Norway}
\author{Sol H. Jacobsen}
\email[Corresponding author: ]{sol.jacobsen@ntnu.no}
\affiliation{Center for Quantum Spintronics, Department of Physics, NTNU, Norwegian University of Science and Technology, NO-7491 Trondheim, Norway}

\begin{abstract}
    We propose that geometric curvature and torsion may be used to probe the quality of an uncompensated antiferromagnetic interface, using the proximity effect. We study a helix of antiferromagnetic wire coupled to a conventional superconductor, and show that a density of states measurement can give information about the quality of an uncompensated interface, crucial for many recently predicted antiferromagnetic proximity effects. Furthermore, we show that geometric curvature alone can result in long-ranged superconducting triplet correlations in the antiferromagnet, and we discuss the impact curvature and torsion can have on the future development of superconducting spintronic devices. 
\end{abstract}

\maketitle

\section{Introduction}
Superconducting spintronics combines the dissipationless spin and charge transport of superconductivity with the information processing capabilities of magnetic heterostructures, reducing energy consumption for some processes, and enabling new ways of performing of computations \cite{Eschrig2011,Linder2015}. With their majority spin-polarized structure addressable with external magnetic fields, ferromagnets (F) have historically been the primary magnetic element in most spintronic architectures. Antiferromagnets (AF), with a lattice of alternating spin orientations, are now becoming serious contenders to replace or supplement traditional ferromagnetic elements due to their robustness in magnetic fields, lack of stray fields, and fast terahertz dynamics \cite{Jungwirth2016,Baltz2018,Xiong2022}. A plethora of recent studies have emphasised the importance of interface characteristics when using antiferromagnetic elements in superconducting spintronic structures, and in particular the importance of a finite interface magnetization via an uncompensated interface \cite{Andersen2005,Kamra2018,Erlandsen2019,Erlandsen2020,Thingstad2021,Bobkov2021,Fyhn2022}. However, manufacturing and characterizing the quality of proximity-coupled uncompensated interfaces is extremely experimentally challenging. In this article, we show that geometric curvature can be used to probe the interface characteristics, and control proximity effects in superconductor-antiferromagnet structures. 

AF terahertz dynamics can enable ultra fast information processing and storage, e.g. in writing \cite{Olejnik2018}, and driving spin-lattice coupling \cite{Mashkovich2021}, which can for example induce emission of terahertz coherent magnons \cite{Rongione2023}. However, a standing issue in manipulating any magnetic element in spintronics is a limited number of external control mechanisms. Some manipulation of the AF order parameter has been shown in experiment \cite{Park2011,Loth2012,Marti2014,Wadley2016}, including isothermal AF switching with pulsed gate voltages, enabling all-electrical readout in AF-based magnetic random access memories (MRAM) \cite{Kosub2017}. In AF spin valves and tunnel junctions, the spin transfer torque can manipulate the AF order parameter \cite{Nunez2006,Merodio2014}, potentially switching the N\'eel vector and representing the writing operation in AF-MRAMs, and the direction of spins at an uncompensated edge can be manipulated using a spin-orbit torque \cite{Lin2019,Liu2020,Zhang2021}. Moreover, exchange-biased bilayers composed by F and uncompensated AF layers, have been used as a platform to investigate the interface magnetization of the AF layer \cite{Roy2005}, map its motion \cite{Zhou2015} and probe its temperature and magnetic field dependence \cite{Lapa2020}.

Geometric curvature has recently emerged as a novel way to design and control spin dynamics using real space manipulation, with a range of interesting and unexpected phenomena at the nanoscale \cite{Gentile2022,Streubel2016,Streubel2021, Gentile2013,Gentile2015,Ortix2010,Ortix2011,Francica2020,Volkov2019,Frustaglia2020,Rodriguez2021,Kutlin2020,Chou2021,Wang2019,Nagasawa2013,Ying2020}. For instance, geometric curvature can promote topological superconductivity in Rashba nanowires \cite{Francica2020} and 2D topological insulators \cite{Chou2021}, and allows for independent geometrical control of spin and charge resistances \cite{Das2019}, as well as control of the spin phase of electrons \cite{Nagasawa2013,Frustaglia2020,Ying2020,Rodriguez2021}. Geometric curvature in a magnet provides an effective spin-orbit coupling due to a changing spin quantization axis \cite{Salamone2021,Salamone2022}, and is addressable \textit{in situ} via for example dynamical strain, photostriction, piezoelectrics, thermoelectrics or tuning the surface chemistry \cite{Kundys2015,Matzen2019,Guillemeney2022}, broadening the experimental toolbox for magnetic order manipulation in spintronic structures. The curvature in a ferromagnetic weak link of a Josephson junction can directly control the direction of current flow through the junction \cite{Salamone2021}, and can even control the superconducting transition, creating a large spin-valve effect \cite{Salamone2022}.

\begin{figure}
    \centering
    \includegraphics[width=\columnwidth]{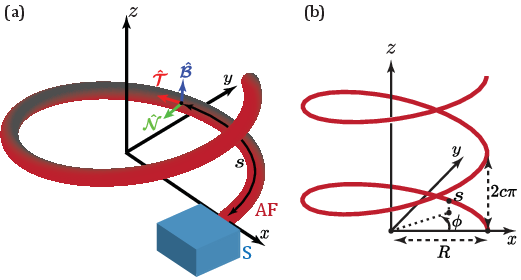}
    \caption{(a) Bulk superconductor coupled to antiferromagnetic helical nanowire, showing the orthonormal unit vectors $\T$, $\N$ and $\B$ parametrizing the curvilinear coordinate system. (b) Characteristic parameters of a helix, with radius $R$, arclength $s$, pitch $2c\pi$, and azimuthal angle $\phi$.}
    \label{fig:SF_helix}
\end{figure}

Manufacturing curved nanostructures has come a long way from the first examples of using chemical etching to roll up nanotubes in the early 2000s \cite{Prinz2000,schmidt2001}. Since then, many new processes and fabrication techniques allow for a range of intricately curved geometries at the nanoscale, for instance via electron-beam lithography \cite{Volkov2019}, two-photon lithography \cite{Sahoo2018}, glancing angle deposition \cite{Gibbs2014} and focused-electron beam induced deposition \cite{Dobrovolskiy2021,Sanz-Hernandez2020,Skoric2020}. While for curved ferromagnets there have been plenty of experimental and theoretical investigations, the field of curvilinear antiferromagnets is still in its infancy, with only a few recent studies of curvature in AF spin chains \cite{CastilloSepulveda2017,Pylypovskyi2020,Yershov2022,Borysenko2022,Makarov2022}. It has been shown that a curvilinear AF helix has chiral helimagnetic behaviour, and that the geometric curvature and torsion can control both the orientation of the N\'eel vector in the ground state and the strength of the Dzyaloshinskii–Moriya interaction, resulting in the hybridization of spin wave modes \cite{Pylypovskyi2020}. In this article, we will consider such a AF helix, proximity coupled to a bulk, conventional superconductor, as illustrated in Fig.~\ref{fig:SF_helix}.

The proximity effect in diffusive superconductor-ferromagnet heterostructures allows the conversion of s-wave singlet superconducting correlations, which rapidly decohere in the F, into so-called long-ranged, spin-polarized triplet correlations, which persist in longer magnetic samples. This conversion can be achieved through the presence of magnetic inhomogeneities \cite{Bergeret2001,Khaire2010,Robinson2010}, spin-orbit coupling \cite{Bergeret2013,Bergeret2014}, or geometric curvature \cite{Salamone2021,Salamone2022}. Theoretical studies combining superconductors with AFs have typically considered clean systems \cite{Bulaevskii2017,Jakobsen2020,Jakobsen2021,Johnsen2021} or modeled the AF as a normal metal \cite{Hauser1966,Zhen2019,Hubener2002}, but very recently a theory for incorporating AF insulators \cite{Bobkov2022} and AF metals (AFMs) \cite{Fyhn2022, Fyhn2023} into the quasiclassical theory for diffusive transport has been developed. In this paper, we will use this quasiclassical approach, in which the AFM behaves like as a normal metal with magnetic impurities, to derive the diffusive transport equations for curved AFM wires proximity coupled to bulk conventional superconductors. We will show that the curvature and torsion alone control the conversion between singlet and long-ranged triplet superconducting correlations, giving a long-ranged proximity effect, as well as a large qualitative change in the local density of states from gap to peak, which can be used to characterize the quality of the uncompensated interface.

The article is organized as follows. In \cref{sec:diff_AF} we briefly discuss the newly developed quasiclassical theory for AFMs \cite{Fyhn2023}. We then generalize this theory in \cref{sec:diff_CurvedAF}, to include geometric curvature in a proximity coupled antiferromagnet. In \cref{sec:Results} we discuss analytic limits of the equations, and present numerical results for the density of states and long range spin correlation functions in a superconducting-antiferromagnetic helix hybrid nanowire. We conclude in \cref{sec:Discussion} with a summary of the results and a discussion of the impact of geometric curvature on the future development of superconducting spintronic devices. 

\section{Diffusive Theory of Antiferromagnetic Metals} \label{sec:diff_AF}
The general approach used to derive the quasiclassical theory in normal metals, superconductors and ferromagnetic metals is modified when treating antiferromagnetic metals, to account for their alternating magnetization on the two different sublattices. Recently, a quasiclassical theory for AFMs in the dirty limit has been developed in Ref. \cite{Fyhn2023}, where they use a square lattice with two sublattices, one for each spin direction. The AFM term enters the Usadel equation in a similar way to magnetic spin-flip impurities in normal metals. This is because the impurities in an antiferromagnet behave as if they were magnetic due to the hybridized spin-sublattice conduction states coupling differently to impurities on the two sublattices ~\cite{Fyhn2023}. This theory has been applied to the study of the proximity effect in a superconductor-AFM hybrid structure, showing the appearance of long-ranged triplets \cite{Fyhn2022}. We note that a quasiclassical theory for AF insulators with superconductivity has also been developed recently \cite{Bobkov2022}. The approaches in Refs.~\cite{Fyhn2023,Fyhn2022} and Ref.~\cite{Bobkov2022} differ in that they assume different sizes for the chemical potential. In the former it is assumed large, meaning that only two of the four AFM bands contribute significantly to the dynamics of the system. Therefore, only two AFM bands are included in the quasiclassical theory, resulting in a Usadel equation with the same matrix structure as for ferromagnetic metals. By employing this model for our system, the role of the chemical potential is simply to control the effective impurity strength. In the latter approach \cite{Bobkov2022}, the chemical potential is assumed to be close to zero. This allows for interband effects, necessitating the inclusion of all four AF bands. The difference between the two cases, as well as the crossover between them, is discussed in Ref.~\cite{Bobkov2023}. 

Below we introduce the Usadel equation for straight antiferromagnets, before we extend the quasiclassical theory of AFMs developed in Ref.~\cite{Fyhn2023} to include geometric curvature in Sec.~\ref{sec:diff_CurvedAF}.

\subsection{Usadel Equation for Straight Antiferromagnets}
The Usadel equation for a time independent superconductor-AFM hybrid structure, with N\'eel vector parallel to the $z$ axis, is \cite{Fyhn2023}
\begin{equation}
    i \Tilde{\nabla}\hat{\vb{j}} + \comm{\hat{\rho}_3\varepsilon - \hat{V}_s + \frac{iJ^2}{2 \tau_{\mathrm{imp}}\mu^2}\hat{\sigma}_z\gs \hat{\sigma}_z}{\gs} = 0, \label{eq:AF_UsadelFull}
\end{equation}
where $\hat{\vb{j}}$ is the matrix current
\begin{equation}
    \hat{\vb{j}} = -\gs\Tilde{\nabla}(D \gs) - \gs\comm{\frac{J^2}{2\mu^2}\hat{\sigma}_z\gs\hat{\sigma}_z}{\hat{\vb{j}}}. \label{eq:AF_matrixcurr}
\end{equation}
Here, $\gs$ is the retarded, isotropic quasiclassical Green's function, a $4\times4$ matrix in Nambu$\times$spin-space, and $\Tilde{\nabla}$ is the covariant derivative defined as $\Tilde{\nabla}\hat{g}=\nabla\hat{g}-i\comm{\hat{\bm{A}}}{\hat{g}}$ where $\hat{\bm{A}}=\diag(\bm{A}, -\bm{A}^*)$ with $\bm{A}$ being a vector field due to electromagnetic fields or spin-orbit coupling. $\hat{\rho}_3=\mathrm{diag}(1,1,-1,-1)$ and $\hat{\sigma}_z=\mathrm{diag}(\sigma_z,\sigma_z^*)$ are matrices in Nambu$\times$spin-space, while $\sigma_z$ is the third Pauli-matrix in spin-space. We note that the complex conjugation in $\hat{\sigma}_z$ has no effect on the third Pauli matrix, but we include it in the notation to maintain notational consistency with other matrices defined below. Furthermore, $\varepsilon$ is the energy, $\hat{V}_s$ the isotropic part of the various potential terms, such as the superconducting pair potential, $\tau_{\mathrm{imp}}$ the elastic impurity scattering time, and $J$ is the exchange energy. The parameter $\mu$ is the chemical potential \cite{Fyhn2022}. 

The Green's function is found by writing it in the band basis, and then using the Green's function associated with the energy band which crosses the Fermi surface \cite{Fyhn2023}.

In \cref{eq:AF_UsadelFull,eq:AF_matrixcurr} we have chosen a different basis from Ref.\cite{Fyhn2023}, where the basis vector was defined with a minus sign multiplying the hole operator with spin up, while in our basis choice the minus sign is absent and the order of the spin up and spin down hole operators is inverted. To go from our form of \cref{eq:AF_UsadelFull,eq:AF_matrixcurr} to those presented in Ref.\cite{Fyhn2023}, one needs to substitute $\hat{\sigma}_z\rightarrow\hat{\sigma}_z\hat{\rho}_3$. Explicitly, our basis choice reads

\be
    \psi^\dagger=(c^\dagger_\uparrow,c^\dagger_\downarrow,c_\uparrow,c_\downarrow),
    \label{eq:basis_choice}
\ee

\noindent where the operator $c^{(\dagger)}_\sigma$, with $\sigma=\uparrow,\downarrow$, annihilates (creates) an electron with spin $\sigma$.

A simpler expression for the matrix current can be found if one assumes that $(J/\mu)^2[\hat{\sigma}_z, \gs] \ll 1$, which is valid in the limit of small $J/\mu$ or vanishing $[\hat{\sigma}_z, \gs]$. The matrix current then becomes \cite{Fyhn2023}

\begin{equation}
    \hat{\vb{j}} \approx -[1 + (J/\mu)^2]^{-1} D \gs\Tilde{\nabla}\gs.
\end{equation}
For a system in equilibrium, the approximated Usadel equation becomes

\begin{equation} \label{eq:AF Usadel}
    i\tilde{D}\Tilde{\nabla} (\gs \Tilde{\nabla} \gs) =
    \comm{\rho_3\varepsilon - \hat{V}_s + \frac{iJ^2}{2 \tau_{\mathrm{imp}}\mu^2} \hat{\sigma}_z\gs\hat{\sigma}_z}{\gs},
\end{equation}
where $\tilde{D} = D\left[ 1 + (J/\mu)^2 \right]^{-1}$ is the renormalized diffusion constant.

\section{Geometric curvature in the diffusive transport theory for antiferromagnetic metals} \label{sec:diff_CurvedAF}
We extend the Usadel \cref{eq:AF Usadel} to apply to curvilinear antiferromagnets by transforming into the Frenet-Serret frame in \cref{sec:curvedUsadel}. We discuss compensated and uncompensated interfaces, and introduce boundary conditions in \cref{sec:BC}. We paramaterize the equations in Riccati form in \cref{sec:curvedRiccati}, in order to later solve the equations numerically.

\subsection{Usadel Equation for Curved Antiferromagnets in Frenet-Serret frame}\label{sec:curvedUsadel}
The three dimensional space around the helix in \cref{fig:SF_helix} can be parameterized as $\vb{R}(s, n, b) = \vb{r}(s) + \curvba{N}(s) n + \curvba{B}(s) b$, where $\vb{r}(s)$ is the parametrization of the curve along the arclength $s$, and $n$ and $b$ are the normal and binormal coordinates, respectively. The geometry of the system is therefore determined by the set of orthogonal unit vectors $\curvba{T}(s) = \partial_s \vb{r}(s)$, $\curvba{N}(s) = \partial_s \curvba{T}(s)/\left|\del_s\T(s)\right|$ and $\curvba{B}(s) = \curvba{T}(s) \times \curvba{N}(s)$ in the tangential, normal and binormal curvilinear directions, respectively. These vectors allow to identify the curvature and torsion of the structure as $\kappa(s)=|\del_s\T(s)|$ and $\tau(s)=|\del_s\B(s)|$, respectively, and are connected through the Frenet-Serret formulas \cite{ortix2015quantum}

\begin{equation}
    \begin{pmatrix} \partial_s \curvba{T}(s) \\ \partial_s \curvba{N}(s) \\ \partial_s \curvba{B}(s)  \end{pmatrix} = \begin{pmatrix} 0 & \kappa(s) & 0 \\ -\kappa(s) & 0 & \tau(s) \\ 0 & -\tau(s) & 0 \end{pmatrix} \begin{pmatrix} \curvba{T}(s) \\ \curvba{N}(s) \\ \curvba{B}(s) \end{pmatrix},
\end{equation}
The metric tensor of this system is given as

\begin{equation} \label{eq:metric_tensor}
    \mathcal{G}_{\mu \nu} = \begin{pmatrix} \eta(s, n)^2 + \tau(s)^2 (n^2 + b^2) & -\tau(s)b & \tau(s)n \\ -\tau(s)b & 1 & 0 \\ \tau(s)n & 0 & 1 \end{pmatrix},
\end{equation}
where $\eta(s, n) = 1 - \kappa(s) n$.

Using tensor notation, the Usadel equation in \eqref{eq:AF Usadel}, for a general orientation of the N\'eel vector $\bm{n}$ is written as

\begin{equation}\begin{split}
    i\tilde{D}&\mathcal{G}^{\lambda \mu} \tensder_\lambda (\gs \tensder_\mu \gs)\\=
    &\comm{\hat{\rho}_3\varepsilon\!-\!\hat{\Delta}\!+\!\frac{iJ^2}{2 \tau_{\mathrm{imp}}\mu^2} (\bm{n}\!\cdot\!\hat{\bm{\sigma}})\gs (\bm{n}\!\cdot\!\hat{\bm{\sigma}})}{\gs},
    \label{eq:Usadel_covForm}
\end{split}\end{equation}
where $\hat{\bm{\sigma}}=\mathrm{diag}(\bm{\sigma},\bm{\sigma}^*)$ with the Pauli vector expressed in curvilinear coordinates $\bm{\sigma}=(\sigma_T,\sigma_N,\sigma_B)$, and $\hat{\Delta} = \antidiag{(\Delta, -\Delta, \Delta^*, -\Delta^*)}$, with $\Delta$ the superconducting order parameter as the potential, which is zero inside the AFM. We also defined the space-gauge covariant derivative as

\begin{equation}
    \tensder_\lambda v_\mu = \Tilde{\partial}_\lambda v_\mu - \Gamma_{\lambda \mu}^\nu v_\nu,
\end{equation}
where the gauge-only covariant derivative term is defined as $\Tilde{\partial}_\lambda v_\mu = \partial_\lambda v_\mu - i \comm{\hat{A}_\lambda}{v_\mu}$, with $\hat{A}_\lambda = \diag{(A_\lambda, -A_\lambda^*)}$, and $\Gamma_{\lambda\mu}^\nu$ are the Christoffel symbols relating derivatives of basis vectors to the basis vectors themselves at position $\vb{R}$ \cite{Kelly2020,Salamone2022}.

We will now look at the case of a nanowire, taking the limit $n, b \rightarrow 0$. With this simplification it can be shown that all Christoffel symbols are zero, such that the left hand side of \cref{eq:Usadel_covForm} reduces to the same form as in Ref.\cite{Salamone2022}. Therefore, the Usadel equation takes the form 
\begin{equation}\label{eq:1d arc Usadel equation}
    i\tilde{D}\Tilde{\partial}_s (\hat{g} \Tilde{\partial}_s \hat{g})\!=\!\comm{\varepsilon\hat{\rho}_3\!-\!\hat{\Delta}\!+\!\frac{i J^2}{2 \tau_{\mathrm{imp}} \mu^2} (\bm{n}\!\cdot\!\hat{\bm{\sigma}})\hat{g}(\bm{n}\!\cdot\!\hat{\bm{\sigma}})}{\hat{g}}.
\end{equation}
This generalized form includes the possibility for both superconductivity and antiferromagnetism to be simultaneously present. In the following sections, we consider instead an antiferromagnet where superconductivity is included through the proximity effect only, i.e. via the boundary conditions. The effect of the curvature now enters the equations through the Pauli matrices depending on the direction of the N\'eel vector $\bm{n}$.
As shown in Ref.\cite{Pylypovskyi2020} for the case of an antiferromagnetic helix, the N{\'e}el vector orientation of the equilibrium state in the curvilinear coordinate system can either be homogeneous, \ie constant orientation with respect to the curved reference frame, or periodic, \ie varying as a function of the arclength coordinate along the geometry of the structure, depending on the strength of the DMI. In what follows, we will assume the AFM to be in the homogeneous state, and we will consider the N\'eel vector to be directed along the binormal direction, \ie $\bm{n}\equiv\B$.

A helical nanowire like the one depicted in \cref{fig:SF_helix}, having radius $R$ and $2c\pi$ pitch, i.e. the height of a complete helix turn, can be defined in cylindrical coordinates as \cite{ortix2015quantum}:
\bse\begin{align}
	x &= R\cos\phi,\\
	y &= R\sin\phi,\\
	z &= c\phi,
\end{align}\ese

\noindent where $\phi=\left[0,2n\pi\right]$ is the azimuthal angle as shown in \cref{fig:SF_helix}, and $n$ defines the number of turns. The value of $c$ determines how much the nanowire is tilted out of plane. Curvature and torsion are respectively given by 
\bse\begin{align}
\kappa&=R/(R^2+c^2) , \label{eq:kappa_exp}\\
\tau&=c/(R^2+c^2). \label{eq:tau_exp}
\end{align}\ese

\noindent The arclength coordinate is given by $s=\phi\sqrt{R^2+c^2}$. This parametrization leads to the following set of three unit vectors:
\begin{subequations}
\begin{align}
    \curvba{T}(s) &= -\cos\alpha\sin\phi\hat{\basis}_x + \cos\alpha\cos\phi\hat{\basis}_y + \sin\alpha\hat{\basis}_z, \\
    \curvba{N}(s) &= -\cos\phi\hat{\basis}_x - \sin\phi\hat{\basis}_y, \\
    \curvba{B}(s) &= \sin\alpha\sin\phi\hat{\basis}_x - \sin\alpha\cos\phi\hat{\basis}_y + \cos\alpha\hat{\basis}_z,
\end{align}
\end{subequations}
where $\alpha = \arctan{(\tau/\kappa)}$. Having derived the form of the three curvilinear unit vectors, the curvilinear Pauli matrices are calculated from
\begin{equation} \label{eq:Curvilinear Pauli matrices}
    \sigma_T = \bm{\sigma} \cdot \curvba{T}(s), \quad \sigma_N = \bm{\sigma} \cdot \curvba{N}(s), \quad \sigma_B = \bm{\sigma} \cdot \curvba{B}(s).
\end{equation}
This form of the Pauli matrices allows for introducing the dependence on the geometry of the system in the model.

\subsection{Boundary Conditions -- Compensated and Uncompensated Interfaces}\label{sec:BC}
To include AFMs in hybrid structures, suitable boundary conditions are needed. These have been derived in the diffusive regime in Ref.\cite{Fyhn2023}, for a S-AF interface. The boundary condition for the matrix current going from material $a=\{S,AF\}$ to material $b=\{S,AF\}$ is \cite{Fyhn2022,Fyhn2023}:
\be
    \bm{e}_n\cdot\hat{\bm{j}}_a=[\hat{T}_{ab}\hat{g}_b\hat{T}_{ba}+i\hat{R}_a,\hat{g}_a],
    \label{eq2:bc_AF_sa}
\ee
where $\bm{e}_n$ is the outward unit vector normal to the interface, $\hat{T}_{ab}$ is the tunneling matrix and $\hat{R}_a$ is the reflection matrix. We note that the above equation is valid only for $a\neq b$. This equation coincides with the generalization of the Kupriyanov-Lukichev boundary conditions for spin-active interfaces \cite{Machon2013}. 

In the case of compensated interfaces, $\hat{T}_{ab}=t$ and $\hat{R}_a$ are scalars and the boundary conditions of \cref{eq2:bc_AF_sa} reduce to the usual Kupriyanov-Lukichev boundary conditions \cite{KuprianovLukichev1988}. For uncompensated interfaces, assuming that the tunneling occurs between the superconductor and one sublattice, the tunneling matrix is \cite{Fyhn2022}:
\be
    \hat{T}_{S,AF}=\frac{1}{2}\left(t_0+t_1\bm{m}\cdot\hat{\bm{\sigma}}\right),
\ee

\noindent where $t_0=t(\sqrt{1+J/\abs{\mu}}+\sqrt{1-J/\abs{\mu}})$ and $t_1=t(\sqrt{1+J/\abs{\mu}}-\sqrt{1-J/\abs{\mu}})$. The unit vector $\bm{m}$ identifies the direction of the interface magnetization. The reflection matrix can be set equal at both sides of the interface, taking the value \cite{Fyhn2022} $\hat{R}_S=\hat{R}_{AF}=G_\varphi\bm{m}\cdot\hat{\bm{\sigma}}$. The factor $G_\varphi$ represents an interfacial phase shift acquired during reflection, like the phase shift picked up at ferromagnetic interfaces \cite{Eschrig2011}, and is a key ingredient in determining the quality of an uncompensated interface, which we will discuss further in \cref{sec:Results}.

\subsection{Riccati Parameterization}\label{sec:curvedRiccati}

To solve the curvilinear Usadel equation with antiferromagnetic coupling \eqref{eq:1d arc Usadel equation} numerically, we use the Riccati parametrization of the quasiclassical Green's function \cite{Jacobsen2015b,Schopohl1995}:
\begin{equation} \label{eq:Riccati parametrization}
    \gR = \begin{pmatrix} N & 0 \\ 0 & -\Tilde{N} \end{pmatrix} \begin{pmatrix} 1 + \gamma \tilde{\gamma} & 2 \gamma \\ 2 \Tilde{\gamma} & 1 + \Tilde{\gamma} \gamma \end{pmatrix},
\end{equation}
where $N$ is a $2 \times 2$ normalization matrix defined as $N = (1 - \gamma \Tilde{\gamma})^{-1}$. The tilde operation is defined as $\Tilde{\gamma}(s, \varepsilon) = \gamma^*(s, -\varepsilon)$. 
Inserting equation \eqref{eq:Riccati parametrization} into equation \eqref{eq:1d arc Usadel equation}, we get the Riccati parameterized Usadel equation of the system:

\begin{equation} \label{eq:Parametrized AF Usadel}
\begin{split}
     &\tilde{D}[(\partial_s^2 \gamma) + 2(\partial_s \gamma)\Tilde{N}\Tilde{\gamma}(\partial_s \gamma)]=-2i\varepsilon\gamma\\
     &+\!2\zeta\Big[\sigma_BN\sigma_B\gamma\!+\!\gamma\sigma^*_B\Tilde{N}\sigma^*_B\!-\!\sigma_BN\gamma\sigma^*_B\!-\!\gamma\sigma^*_B\Tilde{N} \Tilde{\gamma}\sigma_B\gamma\!-\!\gamma\Big],
\end{split}
\end{equation}
where $\zeta \equiv J^2/2\tau_{\mathrm{imp}} \mu^2$ is the effective magnetic impurity strength.

Regarding the boundary conditions, we model the resultant interfacial magnetization of an uncompensated interface via the spin-active boundary conditions of \cref{eq2:bc_AF_sa}. The equation for the AF side of the interface, with the use of the Riccati parametrization of \cref{eq:Riccati parametrization}, can be written as:

\be
    \frac{1}{1+(J/\mu)^2}\del_I\g_{AF}=\frac{t}{4}(1-\g_{AF}\gt_{AF})(I_1\g_{AF}+I_2),
    \label{eq2:bcsa_AF}
\ee

\noindent where the terms $I_1$ and $I_2$ are:
\bse\begin{align}
    I_1=&-(g_{AF}g_S-f_{AF}\tilde{f}_S)(t_0^2+t_0t_1\bm{m}\cdot\bm{\sigma}) \nonumber \\
        &+(t_0^2+t_0t_1\bm{m}\cdot\bm{\sigma})(g_Sg_{AF}-f_S\tilde{f}_{AF}) \nonumber \\
        &-(g_{AF}\bm{m}\cdot\bm{\sigma}g_S-f_{AF}\bm{m}\cdot\bm{\sigma}^*\tilde{f}_S)(t_0t_1+t_1^2\bm{m}\cdot\bm{\sigma}) \nonumber \\
        &+(t_0t_1+t^2_1\bm{m}\cdot\bm{\sigma})(g_S\bm{m}\cdot\bm{\sigma}g_{AF}-f_S\bm{m}\cdot\bm{\sigma}^*\tilde{f}_{AF}) \nonumber \\
        &+iG_\varphi(\bm{m}\cdot\bm{\sigma}g_{AF}-g_{AF}\bm{m}\cdot\bm{\sigma}), \label{eq:sabc_I1}\\
    I_2=&-(g_{AF}f_S-f_{AF}\tilde{g}_S)(t_0^2+t_0t_1\bm{m}\cdot\bm{\sigma}^*) \nonumber \\
        &+(t_0^2+t_0t_1\bm{m}\cdot\bm{\sigma})(g_Sf_{AF}-f_S\tilde{g}_{AF}) \nonumber \\
        &-(g_{AF}\bm{m}\cdot\bm{\sigma}f_S-f_{AF}\bm{m}\cdot\bm{\sigma}^*\tilde{g}_S)(t_0t_1+t_1^2\bm{m}\cdot\bm{\sigma}^*) \nonumber \\
        &+(t_0t_1+t_1^2\bm{m}\cdot\bm{\sigma})(g_S\bm{m}\cdot\bm{\sigma}f_{AF}-f_S\bm{m}\cdot\bm{\sigma}^*\tilde{g}_{AF}) \nonumber \\
        &+iG_\varphi(f_{AF}\bm{m}\cdot\bm{\sigma}^*-\bm{m}\cdot\bm{\sigma}f_{AF}), \label{eq:sabc_I2}
\end{align}\ese
\noindent and $g_a=2N_a-1$, $f_a=2N_a\g_a$ are the $2\times2$ normal and anomalous Green's functions. 

\section{Results} \label{sec:Results}

We consider a hybrid structure, formed by a bulk, conventional s-wave superconductor coupled to an antiferromagnetic helix, as illustrated in \cref{fig:SF_helix}. We solve the Usadel equation \cref{eq:Parametrized AF Usadel} in the antiferromagnet using a bulk solution for the superconductor, given by $\gamma_{BCS}=\sinh\theta/(1+\cosh\theta)i\sigma_y$, with $\theta=\Delta_0/\varepsilon$, and $\Delta_0$ being the bulk gap. We will assume the N\'eel vector to be oriented along the binormal direction and the AF to be uncompensated. In that case, the S/AF interface is described by the spin-active boundary conditions of \cref{eq2:bcsa_AF} with interface magnetization parallel to the binormal direction. We choose the elastic impurity scattering time $\tau_\mathrm{imp}\Delta_0=0.01$ and the interface parameter $t/\sqrt{\Delta_0\xi_S}=2$, with $\xi_S$ superconducting coherence length of the bulk superconductor. 

In the following we will present results for the density of states and pair correlation in the antiferromagnet. We note that when presenting the results we will always consider a fixed length of the antiferromagnetic helix, imposing a one-to-one correspondence between curvature and torsion. This relationship is obtained by considering that the total length of the helix having $n$ number of turns, radius $R$ and pitch $2\pi c$ is given by $L_{AF}=2\pi n\sqrt{R^2+c^2}$. With the use of \cref{eq:kappa_exp,eq:tau_exp} we get $R^2+c^2=1/(\kappa^2+\tau^2)$, and the length can be expressed as $L_{AF}=2\pi n/\sqrt{\kappa^2+\tau^2}$. For instance, if we fix the length and curvature of the helix, the torsion is given by $\tau=\sqrt{(2\pi n/L_{AF})^2-\kappa^2}$, with $\kappa\leq2\pi n/L_{AF}$, giving the one-to-one correspondence between $\kappa$ and $\tau$. For the remainder of this section, we will consider an AF helix with a single turn $n=1$, length $L_{AF}=2\xi_S$, and impurity scattering time $\tau_{\mathrm{imp}}\Delta_0=0.01$.

\subsection{Weak Proximity Equations} \label{sec:wpeeqs}

To understand how curvature affects the AFM, we will first examine the limit of a weak proximity effect. In this limit the components of the $\gamma$-matrix are expected to be small, i.e. $\abs{\gamma_{ij}} \ll 1$, which means we may neglect terms of the order $\mathcal{O}(\gamma^2)$. Therefore, $N \approx 1$, and the anomalous Green's function given in the upper right block of equation \eqref{eq:Riccati parametrization} reduces to $f = 2\gamma$. The anomalous Green's function is then reduced to singlet and triplet components, where the singlet component is described by a scalar function $f_0$, and the triplet components are encapsulated in the so-called $\vb{d}$-vector,
\begin{equation}\label{Eq:dvec}
    f = (f_0 + \vb{d} \cdot \bm{\sigma})i\tau_2,
\end{equation}
where $\vb{d} = (d_T, d_N, d_B)$, and $\tau_2=\mathrm{antidiag}(-i,i)$.

Using the Pauli spin components in \eqref{eq:Curvilinear Pauli matrices}, the anomalous Green's function takes the form
\begin{widetext}
\begin{equation}
    f = \begin{pmatrix} i\cos(\alpha) e^{-i s/L } d_T + e^{-i s/L } d_N - i \sin(\alpha) e^{-i s/L } d_B & f_s + \sin(\alpha) d_T + \cos(\alpha) d_B \\ -f_s + \sin(\alpha) d_T + \cos(\alpha) d_B & i \cos(\alpha) e^{i s/L } d_T - e^{i s/L } d_N - i \sin(\alpha) e^{i s/L } d_B \end{pmatrix},
\end{equation}
\end{widetext}
where $ L = \sqrt{R^2 + c^2} = 1/\sqrt{\kappa^2 + \tau^2}$, with $c$ and $R$ defined in \cref{fig:SF_helix}. Solving equation \eqref{eq:Parametrized AF Usadel} in the weak proximity limit inside the AFM, where $\Delta = 0$, and ignoring any intrinsic spin-orbit coupling, we get four coupled differential equations for the curvilinear components of the $\vb{d}$-vector and singlet $f_0$ :

\bse
\label{eq2:wp_AF}
\begin{align}
    \frac{i\tilde{D}}{2}&(\del^2_sd_T\!-\!2\kappa\del_sd_N)\!=\!\left(\varepsilon\!+\frac{i}{2}\tilde{D}\kappa^2\right)\!d_T\!-\!\frac{i}{2}\tilde{D}\kappa\tau d_B, \label{eq2:wpdT_AF}\\
    \frac{i\tilde{D}}{2}&(\del^2_sd_N\!+\!2\kappa\del_sd_T-2\tau\del_sd_B)\!=\!\l[\varepsilon\!+\frac{i}{2}\tilde{D}(\kappa^2\!+\!\tau^2)\right]\!d_N, \label{eq2:wpdN_AF}\\
    \frac{i\tilde{D}}{2}&(\del^2_sd_B\!+\!2\tau\del_sd_N)\!=\!\left[\varepsilon\!+\!\frac{i}{2}\!\left(\tilde{D}\tau^2\!+\!4\zeta\right)\!\right]\!d_B\!-\!\frac{i}{2}\tilde{D}\kappa\tau d_T, \label{eq2:wpdB_AF}\\
    \frac{i\tilde{D}}{2}&\del^2_sf_0\!=\!(\varepsilon+2i\zeta)f_0. \label{eq2:wpf0_AF}
\end{align}
\ese
We notice first that there is no conversion from singlets to triplets due to geometric curvature terms, as there is in the ferromagnetic case \cite{Salamone2021,Salamone2022}. Instead, the singlets experience an additional decay term due to the effective magnetic impurities, represented by the imaginary energy contribution of $\zeta$, which produces a decay rate proportional to $J^2/\mu^2$. An effective interface magnetization via an uncompensated interface is therefore needed for singlet-triplet conversion. The role of curvature and torsion is to cause spin-precession and spin-relaxation of the triplet components, and it is instructive to study their relationship. Spin-precession is identified by the terms multiplying first derivatives of any triplet component, and is responsible for conversion between the different triplet components. Spin-relaxation is identified by an additional imaginary component of the triplet energy and represents a loss of spin information, with the additional effective impurity term appearing only for the binormal triplet component.

Using the d-vector formalism of (\ref{Eq:dvec}), short-range triplet (SRT) correlations correspond to spins aligning perpendicularly to the N\'eel vector $\bm{n}$ (or parallel with the d-vector), i.e. $\bm{d}_\parallel=\bm{d}\cdot\bm{n}/\l|\bm{n}\r|$, and long-range triplets (LRT) are parallel to the N\'eel vector: $\bm{d}_\perp=\bm{d}\times\bm{n}/\l|\bm{n}\r|$. The weak proximity equations also allow to distinguish between SRTs and LRTs by analyzing the imaginary contributions to the energy for each component, describing spin relaxation. Singlets have a spin relaxation contribution due to the effective magnetic impurities, and binormal triplets have the same contribution as well as an additional one due to the torsion. Therefore, it is clear that the $d_B$ component decays at a rate even higher than the singlets and can be identified as the SRT component. On the other hand, for tangential and normal triplets, the spin relaxation contribution, and thus their decay rate along the helix, depends only on $\kappa$ and $\tau$, so that they can be identified as LRTs depending on the value of curvature and torsion. 

Analyzing the weak proximity effect equations, in combination with the triplet generation happening at the spin-active interface between a superconductor and an uncompensated antiferromagnet, gives insights on the role of $\kappa$ and $\tau$ in this process. At the S/AF interface, the uncompensated antiferromagnet generates triplet correlations with spin along the binormal direction, having a spin-relaxation prefactor $\epsilon^B_r\sim\tau^2/2+2\zeta$, thus having a higher dacay rate compared to the singlet correlations, whose spin-relaxation prefactor is $\epsilon^0_r\sim2\zeta$. 
Looking at \cref{eq2:wpdT_AF,eq2:wpdN_AF,eq2:wpdB_AF}, we see that for $\kappa = 0, \tau > 0$ there is conversion of the binormal triplets only into normal triplets via the first derivative terms, causing precession between the binormal and normal components of the $\vb{d}$-vector. Even though the effective magnetic impurities do not directly cause the normal component to decay, the decay of $d_B$ indirectly leads to decay also of $d_N$.
However, if $\kappa\neq0$, binormal triplets are converted into both tangential and normal components, whose spin relaxation prefactor is $\epsilon_r\sim\kappa^2/2$. Therefore, $\kappa$ allows to generate LRT components and at the same time is responsible for their decay. Moreover, a finite curvature also leads to precession between $d_T$ and $d_N$, reducing the decay indirectly caused by the decay of $d_B$. This means that $\kappa$ needs to be finite but not too big, in order to have a robust presence of LRT in the system.

\subsection{Pair Correlation: Long-ranged triplets from curvature only}
It has recently been shown that a spin-active interface with polarization aligned perpendicular to the Néel vector of a dirty antiferromagnet can induce long ranged correlations~\cite{Fyhn2023}. Here, we show that parallel spin polarizations also generate LRTs, in curved geometries. Such interfaces naturally arises as uncompensated edges of the antiferromagnet. Hence, in the system here proposed, geometric curvature acts as a mediator for LRT conversion, so that no separate interfacial features are required.

\begin{figure}
    \centering
    \includegraphics[width=\columnwidth]{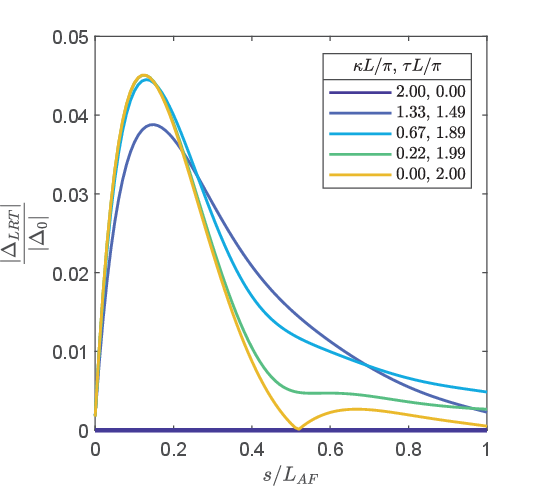}
    \caption{LRT pair correlation as a function of the position in the antiferromagnetic helix for $J^2/\mu^2=0.05$ and different curvature and torsion pairs, with interface parameters $t/\sqrt{\Delta_0\xi_S}=2$, $G_\varphi/\Delta_0\xi_S=1$ and temperature $T=0.005T_{c0}$.}
    \label{fig:LRT_paircorr}
\end{figure}

To see this, consider the singlet and induced triplet pair correlations as a function of the position in the AF, which are defined as
\begin{subequations}
\begin{align}
    \Delta^s(s) ={}& N_0\lambda\int_0^{\Delta_0\cosh(1/N_0\lambda)}\!\!\!d\varepsilon\, f_0(\varepsilon,s)\tanh\!\!\l(\!\frac{\pi}{2e^\gamma}\frac{\varepsilon/\Delta_0}{T/T_{c0}}\!\r),\\
    \Delta^t_\mu(s) ={}& N_0\lambda\int_0^{\Delta_0\cosh(1/N_0\lambda)}\!\!\!d\varepsilon\, d_\mu(\varepsilon,s)\tanh\!\!\l(\!\frac{\pi}{2e^\gamma}\frac{\varepsilon/\Delta_0}{T/T_{c0}}\!\r),
    \label{eq:GapEq}
\end{align}
\end{subequations}
\noindent where $\Delta^s$ is the singlet pair correlation and $\Delta^t_\mu$ is the triplet pair correlation with spin along $\mu$, for $\mu=T,N,B$. In the above equations, $\lambda$ is the coupling constant between electrons, $N_0$ is the density of states at the Fermi level, $\gamma\simeq0.577$ is the Euler-Mascheroni constant, and $T$ is the temperature. $\Delta_0$ and $T_{c0}$ are the superconducting gap and critical temperature of the bulk superconductor, respectively. Considering a conventional s-wave superconductor, we choose the material parameter $N_0\lambda=0.2$.

We demonstrate the presence of LRT correlations by first grouping the long ranged components in a combined pair correlation order parameter:

\be
    \Delta_{LRT}=\sqrt{(\Delta^t_T)^2+(\Delta^t_N)^2}.
\ee

\begin{figure}
    \centering
    \includegraphics[width=\columnwidth]{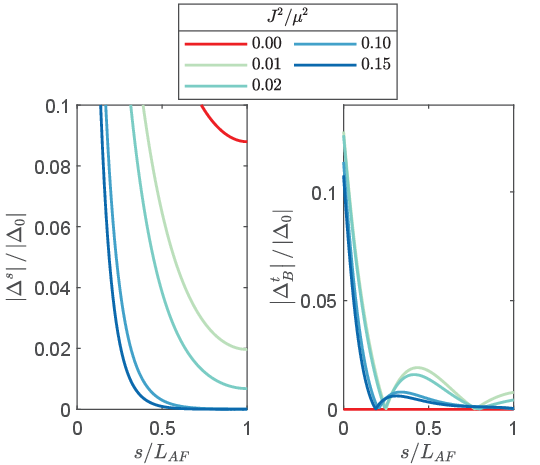}
    \caption{Singlet (left) and binormal (right) pair correlation as a function of the position in the antiferromagnetic helix for $\kappa L_{AF}/\pi\approx0.7$, $\tau L_{AF}/\pi\approx1.9$ and different values of $J^2/\mu^2$, with interface parameters $t/\sqrt{\Delta_0\xi_S}=2$, $G_\varphi/\Delta_0\xi_S=1$ and temperature $T=0.005T_{c0}$. \iffalse The value of the LRT pair correlation at the vacuum interface for $J^2/\mu^2=0.05$ and $\kappa L_{AF}/\pi\approx0.7$, $\tau L_{AF}/\pi\approx1.9$ is plotted with a black dashed line.\fi}
    \label{fig:0B_paircorr}
\end{figure}

\noindent In \cref{fig:LRT_paircorr} we plot the LRT pair correlation as a function of the position in the AF, for $J^2/\mu^2=0.05$, temperature $T=0.005T_{c0}$ and various $\kappa,\tau$ pairs. As expected from the argument presented in \cref{sec:wpeeqs}, for $\kappa=0$ the LRT pair correlation is zero, since there is no SRT-LRT conversion in this case. When $\kappa\neq0$, LRT correlations appear in the AF  with a maximum in the conversion close to the S/AF interface, since the SRTs needed for their conversion decay exponentially away from the interface. After this maximum, the correlations decay until they reach a non zero value at the vacuum interface. This value is higher in the case $\kappa L_{AF}/\pi\sim0.7$ with respect to the other cases; we explore the system dependence on $\kappa$ and $\tau$ further in the following section. For $\kappa=0$ and $\tau L_{AF}/\pi = 2$ the binormal component $d_B$ couples only to the normal component $d_N$ via a spin-precession term, as seen from \cref{eq2:wp_AF}, while the tangential component $d_T$ is completely uncoupled and therefore always zero. Hence, the LRT pair correlation oscillates due to the spin rotation between $d_B$ and $d_N$ in this case. This precession is visible also in the SRT pair correlation (\cref{fig:0B_paircorr}).

To confirm that singlet and SRT pair correlations decay completely and that LRT dominate, we plot singlet and SRT (binormal) pair correlations  as a function of the position in the AF in \cref{fig:0B_paircorr}, taking $\kappa L_{AF}/\pi\approx0.7$, $\tau L_{AF}/\pi\approx1.9$ and different values of $J^2/\mu^2$. We note that the singlet pair correlations do not decay completely at the vacuum edge for $J^2/\mu^2\leq0.02$, while for $J^2/\mu^2>0.02$ they completely die off at about halfway inside the AF. The SRT correlations do not decay as fast as the singlets for $J^2/\mu^2>0.02$. This can be explained by observing that curvature and torsion cause a small degree of conversion from LRT back to SRT, slightly increasing the characteristic decay length for the SRT correlations. However, comparing \cref{fig:0B_paircorr,fig:LRT_paircorr}, it is clear that the SRT decay length is significantly smaller than LRT decay length, confirming LRT dominance. 

\subsection{Density of States and curvature as an interface probe}

The normalized density of states $N(\varepsilon)$ can be expressed in terms of the Riccati matrices as \cite{Jacobsen2015b}:
\be
    N(\varepsilon)=\frac{1}{2}\mathrm{Tr}\l\{N(1+\g\gt)\right\}.
\ee

\noindent In the weak proximity limit the zero energy term $N(0)$ can be written in terms of singlet and triplet components:

\be
N(0) = 1-\frac{\l|f_0\right|^2}{2}+\frac{1}{2}\sum_i\l|d_i\right|^2,\label{Eqn:WPRDOS}
\ee

\noindent with $i=\l\{T,N,B\r\}$. This expression shows that singlets ($f_0$) contribute to lowering the density of states from the normal-state value of one at zero energy, while triplets ($d_i$) contribute to increasing it. Therefore, the appearance of a gap in $N(0)$ is a signature of a singlet-dominated regime and a zero energy peak is a signature of a triplet-dominated regime \cite{Tanaka2007,Jacobsen2015b}.

\begin{figure}
    \centering
    \includegraphics[width=\columnwidth]{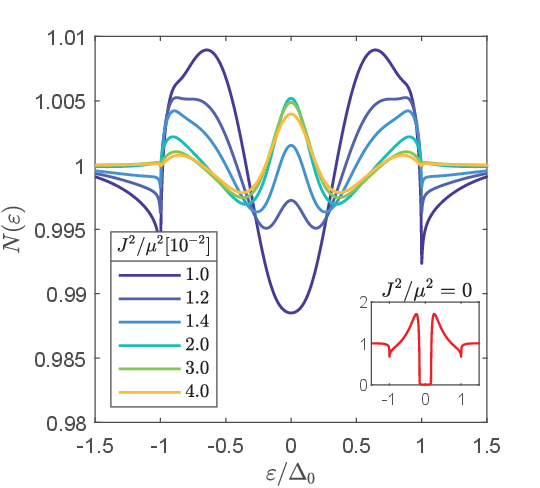}
    \caption{Density of states as a function of the energy at the vacuum edge of the antiferromagnetic helix, for different values of $J^2/\mu^2$ and a fixed curvature and torsion pair $\kappa L_{AF}/\pi\approx0.9$, $\tau L_{AF}/\pi\approx1.8$ with interface parameters $t/\sqrt{\Delta_0\xi_S}=2$, $G_\varphi/\Delta_0\xi_S=1$. The case of $J^2/\mu^2=0$ is plotted in the inset in red.}
    \label{fig:DOS_SAF_n1}
\end{figure}

\begin{figure}
    \centering
    \includegraphics{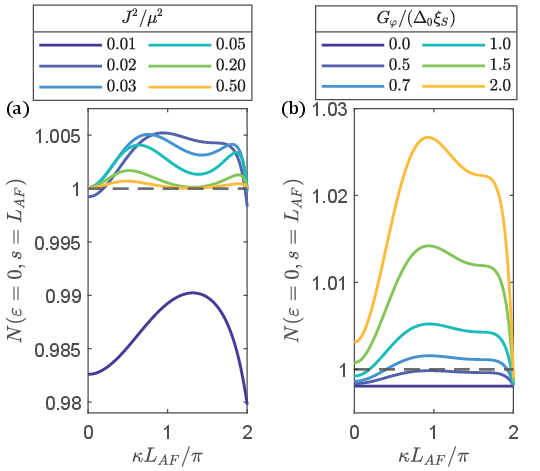}
    \caption{Density of states at zero energy, at the vacuum interface of the antiferromagnetic helix as a function of the curvature $\kappa$ (a) for different values of $J^2/\mu^2$ with $G_\varphi/\Delta_0\xi_S=1$, and (b) for different values of $G_\varphi/\Delta_0\xi_S$ with $J^2/\mu^2=0.02$. In each case, the interface parameter is $t/\sqrt{\Delta_0\xi_S}=2$.}
    \label{fig:dos0E_end_vs_k}
\end{figure}

In Fig.~\ref{fig:DOS_SAF_n1}, we plot the density of states $N(\varepsilon)$ at the vacuum edge of an antiferromagnetic helix with an uncompensated interface at the superconducting edge. We choose curvature $\kappa L_{AF}/\pi\approx0.9$ and torsion $\tau L_{AF}/\pi\approx1.8$, motivated by their roles in the weak proximity equations, in combination with the triplet generation taking place at the spin-active, uncompensated interface. As discussed at the end of \cref{sec:wpeeqs}, both $\kappa$ and $\tau$ lead to generation of LRT components, but their effect differ in that $\kappa$ and $\tau$ lead to the decay of mostly the LRT or SRT, respectively. Hence, the triplets can dominate over the singlets for a wide range of $\kappa$ and $\tau$ values, but with varying relative strength between the LRT and SRT components. This is clear from \cref{fig:dos0E_end_vs_k}, where we explore the importance of the curvature $\kappa$ for the zero energy features at the vacuum edge of the AF wire, for different values of $J^2/\mu^2$. Given the one-to-one correspondence with the torsion $\tau$ when keeping the length fixed, it can be also seen as a function of $\tau$.

The density of states is plotted for different values of the antiferromagnetic exchange $J$ in \cref{fig:DOS_SAF_n1}. In the inset of the figure we plot the case of $J=0$ corresponding to the normal metal case, showing that the system presents a minigap even away from the interface due to the absence of triplet correlations. In contrast, when $J\neq0$ we note that the gap progressively closes when increasing $J$, and at the same time a peak in the density of states at zero energy starts appearing. For $J^2/\mu^2=0.02$, we note a fully formed peak, a clear signal of the presence of triplet correlations in the system (see \cref{Eqn:WPRDOS}). When $J$ is further increased the value of the peak slowly decreases, due to the increased decay of SRT correlations. Zero energy peaks of this order are easily detectable in differential conductance measurements \cite{Kontos2001}.

In \cref{fig:dos0E_end_vs_k}(a) we note that when $J^2/\mu^2=0.01$, varying the curvature does not remove the singlet-induced suppression of the density of states. In contrast, for $J^2/\mu^2=0.02$ increasing the curvature from 0 produces the appearance of a peak in the density of states with a maximum reached for $\kappa L_{AF}/\pi\simeq1$. For higher values of $J^2/\mu^2$ we note a similar behavior, but with progressively smaller peaks, until for $J^2/\mu^2\geq0.2$ and beyond the density of states tends to that of a normal metal with unitary value. This is due to the decay of the SRT (and singlet) correlations caused by the antiferromagnetic exchange through the effective magnetic impurities. When the antiferromagnetic exchange is too high the SRT correlations decay in a very short distance, not long enough for them to feel the effects of geometric curvature and be converted into LRT correlations.

The density of states can now be used to probe the quality of a weakly uncompensated interface. We suggest that the end of the antiferromagnetic wire is the most accessible position for a differential conductance measurement, but the signal strength is greater closer to the superconducting edge. When measuring the density of states, the presence of a minigap would be a signature of a compensated interface, or infinitesimal antiferromagnetic exchange ordering, while a dip or a peak at zero energy would correspond to an uncompensated interface. A higher peak signals greater triplet conversion, and corresponds to a higher quality of the uncompensated interface. When the exchange interaction is very high, the normal metal behaviour dominates and we get no triplet survival at the end of the wire. 

The results in \cref{fig:dos0E_end_vs_k}(a) indicate optimal curvatures for detecting zero-bias peaks in the density of states for different values of the exchange interaction, which gives a useful measure of the efficiency with which the interface converts singlets to triplets. However, this does not directly probe the phase $G_\varphi$ accumulated at the interface due to the uncompensated spins. In \cref{fig:dos0E_end_vs_k}(b), we show that, although the strength of $G_\varphi$ also affects the magnitude of the zero-energy enhancement, the geometric profile remains qualitatively the same. That is, the curvature-controlled maximal peak occurs for the same parameter range, indicating that the same $\kappa$, $\tau$ values will be optimal for probing the degree of compensation at the interface, regardless of the phase picked up due to the magnetization. In experiments, $G_\varphi$ is generally treated as a fitting parameter, and we can see from \cref{fig:dos0E_end_vs_k}(b) that it can greatly enhance the detectable signal. However, since an uncompensated edge is a spin-polarized monolayer at the interface, it is to be expected that its value is rather small. Suggestions, such as the one presented here, for how to probe proximity-coupled uncompensated edges are therefore important in establishing the feasibility of using such edges in spintronic devices.

\section{Discussion} \label{sec:Discussion}
We have developed a quasiclassical theory for diffusive transport in antiferromagnetic helices, using curvilinear coordinates, and study proximity effects when this is coupled to a conventional superconductor. For a conventional superconductor coupled to a straight antiferromagnet, it has been theorized that a misaligned spin-active interface can induce long-ranged triplet superconducting correlations in the antiferromagnet \cite{Fyhn2023}. In that case, there would need to be an intrinsic inhomogeneity, such as a domain wall, at the interface, or an additional layer between the AF and superconductor with a magnetic misalignment to the AF, which can both be problematic, just as in the case for superconductor-ferromagnet systems. In contrast, we find that when the interface is uncompensated, geometric curvature alone can induce long-ranged triplet superconducting correlations in the antiferromagnet , with a decay length tunable by the curvature-torsion relationship, thereby eliminating the need for any additional materials or intrinsic structures.

In general, curvilinear magnetism offers a large, non-relativistic origin of effective spin-orbit coupling, which can be designed and altered within a single sample, and has considerable potential to be harnessed in superconducting spintronics. Real-space misalignment can be controlled using a range of new tools not yet explored in this context. In this article, we considered a homogeneous AF ground state, and it would be interesting to explore the periodic alternative \cite{Pylypovskyi2020}.

To the best of our knowledge there are not yet experimental proposals for curvilinear antiferromagnets, although the prospect has been highlighted as a promising one for future development in curvilinear magnetism reviews (see \eg \cite{Makarov2022}). Curvilinear AFM spin-chain helices have been studied in Ref.~\cite{Pylypovskyi2020}, with the suggestion that copper-based and metal-organic materials may be suitable for testing predictions in that limit. In Ref.~\cite{Zhang2016}, the value of $J$ is estimated to be $|J|=0.8$ cm$^{-1}$ for copper-based materials, while Ref.~\cite{Samanta2014} estimates $|J|$ to be in the range $(0.39-15.11)$ cm$^{-1}$ for antiferromagnetic materials in the DNA metal-organic category. In our work, only the ratio $J/\mu$ enters the calculations, with neither $J$ nor $\mu$ being set explicitly. The calculations should therefore be valid for antiferromagnetic metals with ratios between the antiferromagnetic coupling and chemical potential comparable to what is used here, provided that $J$ and $\mu$ are both sufficiently larger than all other energy scales in the system, c.f. \cite{Fyhn2023}. 

The finite magnetization of an uncompensated antiferromagnetic interface converts singlets to non-polarized triplet pairs, and these are rotated into spin-polarized pairs in the antiferromagnet in a process combining spin precession and diffusion as determined by the relationship between the curvature $\kappa$ and the torsion $\tau$. For a broad range of parameters, there can be considerable triplet conversion, which in turn governs the behaviour of the density of states. We show that the density of states changes from a dip to a peak as a function of the antiferromagnetic exchange interaction, and we therefore propose that a differential conductance measurement at the end of the antiferromagnetic wire can be used to quantify the quality of the uncompensated interface, which is otherwise difficult to characterize in multi-layer heterostructures. We provide analysis of the relationship between the curvature-torsion parameters for different antiferromagnetic exchange strengths. For implementing this probe, static helices can be manufactured near the maximal peak, or the curvature variations can be probed \textit{in situ} by the application of strain. The strength of signal achievable in experiments will ultimately determine the feasibility of the use of uncompensated interfaces in superconducting spintronics.

\begin{acknowledgments}
Computations have been performed on the SAGA supercomputer provided by UNINETT Sigma2 - the National Infrastructure for High Performance Computing and Data Storage in Norway. We acknowledge funding via the “Outstanding Academic Fellows” programme at NTNU, the Research Council of Norway Grant No. 302315, as well as through its Centres of Excellence funding scheme, Project No. 262633, “QuSpin.” MA acknowledges support from the Swedish Research Council (Grant No. VR 2019-04735 of Vladimir Juri\v{c}i\'c). Nordita is supported in part by NordForsk. We thank J.Linder for helpful discussion.
\end{acknowledgments}


\begin{thebibliography}{85}%
\makeatletter
\providecommand \@ifxundefined [1]{%
 \@ifx{#1\undefined}
}%
\providecommand \@ifnum [1]{%
 \ifnum #1\expandafter \@firstoftwo
 \else \expandafter \@secondoftwo
 \fi
}%
\providecommand \@ifx [1]{%
 \ifx #1\expandafter \@firstoftwo
 \else \expandafter \@secondoftwo
 \fi
}%
\providecommand \natexlab [1]{#1}%
\providecommand \enquote  [1]{``#1''}%
\providecommand \bibnamefont  [1]{#1}%
\providecommand \bibfnamefont [1]{#1}%
\providecommand \citenamefont [1]{#1}%
\providecommand \href@noop [0]{\@secondoftwo}%
\providecommand \href [0]{\begingroup \@sanitize@url \@href}%
\providecommand \@href[1]{\@@startlink{#1}\@@href}%
\providecommand \@@href[1]{\endgroup#1\@@endlink}%
\providecommand \@sanitize@url [0]{\catcode `\\12\catcode `\$12\catcode
  `\&12\catcode `\#12\catcode `\^12\catcode `\_12\catcode `\%12\relax}%
\providecommand \@@startlink[1]{}%
\providecommand \@@endlink[0]{}%
\providecommand \url  [0]{\begingroup\@sanitize@url \@url }%
\providecommand \@url [1]{\endgroup\@href {#1}{\urlprefix }}%
\providecommand \urlprefix  [0]{URL }%
\providecommand \Eprint [0]{\href }%
\providecommand \doibase [0]{https://doi.org/}%
\providecommand \selectlanguage [0]{\@gobble}%
\providecommand \bibinfo  [0]{\@secondoftwo}%
\providecommand \bibfield  [0]{\@secondoftwo}%
\providecommand \translation [1]{[#1]}%
\providecommand \BibitemOpen [0]{}%
\providecommand \bibitemStop [0]{}%
\providecommand \bibitemNoStop [0]{.\EOS\space}%
\providecommand \EOS [0]{\spacefactor3000\relax}%
\providecommand \BibitemShut  [1]{\csname bibitem#1\endcsname}%
\let\auto@bib@innerbib\@empty
\bibitem [{\citenamefont {Eschrig}(2011)}]{Eschrig2011}%
  \BibitemOpen
  \bibfield  {author} {\bibinfo {author} {\bibfnamefont {M.}~\bibnamefont
  {Eschrig}},\ }\href {https://doi.org/10.1063/1.3541944} {\bibfield  {journal}
  {\bibinfo  {journal} {Physics Today}\ }\textbf {\bibinfo {volume} {64}},\
  \bibinfo {pages} {43} (\bibinfo {year} {2011})}\BibitemShut {NoStop}%
\bibitem [{\citenamefont {Linder}\ and\ \citenamefont
  {Robinson}(2015)}]{Linder2015}%
  \BibitemOpen
  \bibfield  {author} {\bibinfo {author} {\bibfnamefont {J.}~\bibnamefont
  {Linder}}\ and\ \bibinfo {author} {\bibfnamefont {J.~W.~A.}\ \bibnamefont
  {Robinson}},\ }\href {https://doi.org/10.1038/nphys3242} {\bibfield
  {journal} {\bibinfo  {journal} {Nature Physics}\ }\textbf {\bibinfo {volume}
  {11}},\ \bibinfo {pages} {307} (\bibinfo {year} {2015})}\BibitemShut
  {NoStop}%
\bibitem [{\citenamefont {Jungwirth}\ \emph {et~al.}(2016)\citenamefont
  {Jungwirth}, \citenamefont {Marti}, \citenamefont {Wadley},\ and\
  \citenamefont {Wunderlich}}]{Jungwirth2016}%
  \BibitemOpen
  \bibfield  {author} {\bibinfo {author} {\bibfnamefont {T.}~\bibnamefont
  {Jungwirth}}, \bibinfo {author} {\bibfnamefont {X.}~\bibnamefont {Marti}},
  \bibinfo {author} {\bibfnamefont {P.}~\bibnamefont {Wadley}},\ and\ \bibinfo
  {author} {\bibfnamefont {J.}~\bibnamefont {Wunderlich}},\ }\href
  {https://doi.org/10.1038/nnano.2016.18} {\bibfield  {journal} {\bibinfo
  {journal} {Nature Nanotechnology}\ }\textbf {\bibinfo {volume} {11}},\
  \bibinfo {pages} {231} (\bibinfo {year} {2016})}\BibitemShut {NoStop}%
\bibitem [{\citenamefont {Baltz}\ \emph {et~al.}(2018)\citenamefont {Baltz},
  \citenamefont {Manchon}, \citenamefont {Tsoi}, \citenamefont {Moriyama},
  \citenamefont {Ono},\ and\ \citenamefont {Tserkovnyak}}]{Baltz2018}%
  \BibitemOpen
  \bibfield  {author} {\bibinfo {author} {\bibfnamefont {V.}~\bibnamefont
  {Baltz}}, \bibinfo {author} {\bibfnamefont {A.}~\bibnamefont {Manchon}},
  \bibinfo {author} {\bibfnamefont {M.}~\bibnamefont {Tsoi}}, \bibinfo {author}
  {\bibfnamefont {T.}~\bibnamefont {Moriyama}}, \bibinfo {author}
  {\bibfnamefont {T.}~\bibnamefont {Ono}},\ and\ \bibinfo {author}
  {\bibfnamefont {Y.}~\bibnamefont {Tserkovnyak}},\ }\href
  {https://doi.org/10.1103/RevModPhys.90.015005} {\bibfield  {journal}
  {\bibinfo  {journal} {Rev. Mod. Phys.}\ }\textbf {\bibinfo {volume} {90}},\
  \bibinfo {pages} {015005} (\bibinfo {year} {2018})}\BibitemShut {NoStop}%
\bibitem [{\citenamefont {Xiong}\ \emph {et~al.}(2022)\citenamefont {Xiong},
  \citenamefont {Jiang}, \citenamefont {Shi}, \citenamefont {Du}, \citenamefont
  {Yao}, \citenamefont {Guo}, \citenamefont {Zhu}, \citenamefont {Cao},
  \citenamefont {Peng}, \citenamefont {Cai}, \citenamefont {Zhu},\ and\
  \citenamefont {Zhao}}]{Xiong2022}%
  \BibitemOpen
  \bibfield  {author} {\bibinfo {author} {\bibfnamefont {D.}~\bibnamefont
  {Xiong}}, \bibinfo {author} {\bibfnamefont {Y.}~\bibnamefont {Jiang}},
  \bibinfo {author} {\bibfnamefont {K.}~\bibnamefont {Shi}}, \bibinfo {author}
  {\bibfnamefont {A.}~\bibnamefont {Du}}, \bibinfo {author} {\bibfnamefont
  {Y.}~\bibnamefont {Yao}}, \bibinfo {author} {\bibfnamefont {Z.}~\bibnamefont
  {Guo}}, \bibinfo {author} {\bibfnamefont {D.}~\bibnamefont {Zhu}}, \bibinfo
  {author} {\bibfnamefont {K.}~\bibnamefont {Cao}}, \bibinfo {author}
  {\bibfnamefont {S.}~\bibnamefont {Peng}}, \bibinfo {author} {\bibfnamefont
  {W.}~\bibnamefont {Cai}}, \bibinfo {author} {\bibfnamefont {D.}~\bibnamefont
  {Zhu}},\ and\ \bibinfo {author} {\bibfnamefont {W.}~\bibnamefont {Zhao}},\
  }\href {https://doi.org/https://doi.org/10.1016/j.fmre.2022.03.016}
  {\bibfield  {journal} {\bibinfo  {journal} {Fundamental Research}\ }\textbf
  {\bibinfo {volume} {2}},\ \bibinfo {pages} {522} (\bibinfo {year}
  {2022})}\BibitemShut {NoStop}%
\bibitem [{\citenamefont {Andersen}\ \emph {et~al.}(2005)\citenamefont
  {Andersen}, \citenamefont {Bobkova}, \citenamefont {Hirschfeld},\ and\
  \citenamefont {Barash}}]{Andersen2005}%
  \BibitemOpen
  \bibfield  {author} {\bibinfo {author} {\bibfnamefont {B.~M.}\ \bibnamefont
  {Andersen}}, \bibinfo {author} {\bibfnamefont {I.~V.}\ \bibnamefont
  {Bobkova}}, \bibinfo {author} {\bibfnamefont {P.~J.}\ \bibnamefont
  {Hirschfeld}},\ and\ \bibinfo {author} {\bibfnamefont {Y.~S.}\ \bibnamefont
  {Barash}},\ }\href {https://doi.org/10.1103/PhysRevB.72.184510} {\bibfield
  {journal} {\bibinfo  {journal} {Phys. Rev. B}\ }\textbf {\bibinfo {volume}
  {72}},\ \bibinfo {pages} {184510} (\bibinfo {year} {2005})}\BibitemShut
  {NoStop}%
\bibitem [{\citenamefont {Kamra}\ \emph {et~al.}(2018)\citenamefont {Kamra},
  \citenamefont {Rezaei},\ and\ \citenamefont {Belzig}}]{Kamra2018}%
  \BibitemOpen
  \bibfield  {author} {\bibinfo {author} {\bibfnamefont {A.}~\bibnamefont
  {Kamra}}, \bibinfo {author} {\bibfnamefont {A.}~\bibnamefont {Rezaei}},\ and\
  \bibinfo {author} {\bibfnamefont {W.}~\bibnamefont {Belzig}},\ }\href
  {https://doi.org/10.1103/PhysRevLett.121.247702} {\bibfield  {journal}
  {\bibinfo  {journal} {Phys. Rev. Lett.}\ }\textbf {\bibinfo {volume} {121}},\
  \bibinfo {pages} {247702} (\bibinfo {year} {2018})}\BibitemShut {NoStop}%
\bibitem [{\citenamefont {Erlandsen}\ \emph {et~al.}(2019)\citenamefont
  {Erlandsen}, \citenamefont {Kamra}, \citenamefont {Brataas},\ and\
  \citenamefont {Sudb\o{}}}]{Erlandsen2019}%
  \BibitemOpen
  \bibfield  {author} {\bibinfo {author} {\bibfnamefont {E.}~\bibnamefont
  {Erlandsen}}, \bibinfo {author} {\bibfnamefont {A.}~\bibnamefont {Kamra}},
  \bibinfo {author} {\bibfnamefont {A.}~\bibnamefont {Brataas}},\ and\ \bibinfo
  {author} {\bibfnamefont {A.}~\bibnamefont {Sudb\o{}}},\ }\href
  {https://doi.org/10.1103/PhysRevB.100.100503} {\bibfield  {journal} {\bibinfo
   {journal} {Phys. Rev. B}\ }\textbf {\bibinfo {volume} {100}},\ \bibinfo
  {pages} {100503(R)} (\bibinfo {year} {2019})}\BibitemShut {NoStop}%
\bibitem [{\citenamefont {Erlandsen}\ \emph {et~al.}(2020)\citenamefont
  {Erlandsen}, \citenamefont {Brataas},\ and\ \citenamefont
  {Sudb\o{}}}]{Erlandsen2020}%
  \BibitemOpen
  \bibfield  {author} {\bibinfo {author} {\bibfnamefont {E.}~\bibnamefont
  {Erlandsen}}, \bibinfo {author} {\bibfnamefont {A.}~\bibnamefont {Brataas}},\
  and\ \bibinfo {author} {\bibfnamefont {A.}~\bibnamefont {Sudb\o{}}},\ }\href
  {https://doi.org/10.1103/PhysRevB.101.094503} {\bibfield  {journal} {\bibinfo
   {journal} {Phys. Rev. B}\ }\textbf {\bibinfo {volume} {101}},\ \bibinfo
  {pages} {094503} (\bibinfo {year} {2020})}\BibitemShut {NoStop}%
\bibitem [{\citenamefont {Thingstad}\ \emph {et~al.}(2021)\citenamefont
  {Thingstad}, \citenamefont {Erlandsen},\ and\ \citenamefont
  {Sudb\o{}}}]{Thingstad2021}%
  \BibitemOpen
  \bibfield  {author} {\bibinfo {author} {\bibfnamefont {E.}~\bibnamefont
  {Thingstad}}, \bibinfo {author} {\bibfnamefont {E.}~\bibnamefont
  {Erlandsen}},\ and\ \bibinfo {author} {\bibfnamefont {A.}~\bibnamefont
  {Sudb\o{}}},\ }\href {https://doi.org/10.1103/PhysRevB.104.014508} {\bibfield
   {journal} {\bibinfo  {journal} {Phys. Rev. B}\ }\textbf {\bibinfo {volume}
  {104}},\ \bibinfo {pages} {014508} (\bibinfo {year} {2021})}\BibitemShut
  {NoStop}%
\bibitem [{\citenamefont {Bobkov}\ \emph {et~al.}(2021)\citenamefont {Bobkov},
  \citenamefont {Bobkova}, \citenamefont {Bobkov},\ and\ \citenamefont
  {Kamra}}]{Bobkov2021}%
  \BibitemOpen
  \bibfield  {author} {\bibinfo {author} {\bibfnamefont {G.~A.}\ \bibnamefont
  {Bobkov}}, \bibinfo {author} {\bibfnamefont {I.~V.}\ \bibnamefont {Bobkova}},
  \bibinfo {author} {\bibfnamefont {A.~M.}\ \bibnamefont {Bobkov}},\ and\
  \bibinfo {author} {\bibfnamefont {A.}~\bibnamefont {Kamra}},\ }\href
  {https://doi.org/10.1103/PhysRevB.103.094506} {\bibfield  {journal} {\bibinfo
   {journal} {Phys. Rev. B}\ }\textbf {\bibinfo {volume} {103}},\ \bibinfo
  {pages} {094506} (\bibinfo {year} {2021})}\BibitemShut {NoStop}%
\bibitem [{\citenamefont {Fyhn}\ \emph
  {et~al.}(2023{\natexlab{a}})\citenamefont {Fyhn}, \citenamefont {Brataas},
  \citenamefont {Qaiumzadeh},\ and\ \citenamefont {Linder}}]{Fyhn2022}%
  \BibitemOpen
  \bibfield  {author} {\bibinfo {author} {\bibfnamefont {E.~H.}\ \bibnamefont
  {Fyhn}}, \bibinfo {author} {\bibfnamefont {A.}~\bibnamefont {Brataas}},
  \bibinfo {author} {\bibfnamefont {A.}~\bibnamefont {Qaiumzadeh}},\ and\
  \bibinfo {author} {\bibfnamefont {J.}~\bibnamefont {Linder}},\ }\href
  {https://doi.org/10.1103/PhysRevLett.131.076001} {\bibfield  {journal}
  {\bibinfo  {journal} {Phys. Rev. Lett.}\ }\textbf {\bibinfo {volume} {131}},\
  \bibinfo {pages} {076001} (\bibinfo {year} {2023}{\natexlab{a}})}\BibitemShut
  {NoStop}%
\bibitem [{\citenamefont {Olejník}\ \emph {et~al.}(2018)\citenamefont
  {Olejník}, \citenamefont {Seifert}, \citenamefont {Kašpar}, \citenamefont
  {Novák}, \citenamefont {Wadley}, \citenamefont {Campion}, \citenamefont
  {Baumgartner}, \citenamefont {Gambardella}, \citenamefont {Němec},
  \citenamefont {Wunderlich}, \citenamefont {Sinova}, \citenamefont {Kužel},
  \citenamefont {Müller}, \citenamefont {Kampfrath},\ and\ \citenamefont
  {Jungwirth}}]{Olejnik2018}%
  \BibitemOpen
  \bibfield  {author} {\bibinfo {author} {\bibfnamefont {K.}~\bibnamefont
  {Olejník}}, \bibinfo {author} {\bibfnamefont {T.}~\bibnamefont {Seifert}},
  \bibinfo {author} {\bibfnamefont {Z.}~\bibnamefont {Kašpar}}, \bibinfo
  {author} {\bibfnamefont {V.}~\bibnamefont {Novák}}, \bibinfo {author}
  {\bibfnamefont {P.}~\bibnamefont {Wadley}}, \bibinfo {author} {\bibfnamefont
  {R.~P.}\ \bibnamefont {Campion}}, \bibinfo {author} {\bibfnamefont
  {M.}~\bibnamefont {Baumgartner}}, \bibinfo {author} {\bibfnamefont
  {P.}~\bibnamefont {Gambardella}}, \bibinfo {author} {\bibfnamefont
  {P.}~\bibnamefont {Němec}}, \bibinfo {author} {\bibfnamefont
  {J.}~\bibnamefont {Wunderlich}}, \bibinfo {author} {\bibfnamefont
  {J.}~\bibnamefont {Sinova}}, \bibinfo {author} {\bibfnamefont
  {P.}~\bibnamefont {Kužel}}, \bibinfo {author} {\bibfnamefont
  {M.}~\bibnamefont {Müller}}, \bibinfo {author} {\bibfnamefont
  {T.}~\bibnamefont {Kampfrath}},\ and\ \bibinfo {author} {\bibfnamefont
  {T.}~\bibnamefont {Jungwirth}},\ }\href
  {https://doi.org/10.1126/sciadv.aar3566} {\bibfield  {journal} {\bibinfo
  {journal} {Science Advances}\ }\textbf {\bibinfo {volume} {4}},\ \bibinfo
  {pages} {eaar3566} (\bibinfo {year} {2018})}\BibitemShut {NoStop}%
\bibitem [{\citenamefont {Mashkovich}\ \emph {et~al.}(2021)\citenamefont
  {Mashkovich}, \citenamefont {Grishunin}, \citenamefont {Dubrovin},
  \citenamefont {Zvezdin}, \citenamefont {Pisarev},\ and\ \citenamefont
  {Kimel}}]{Mashkovich2021}%
  \BibitemOpen
  \bibfield  {author} {\bibinfo {author} {\bibfnamefont {E.~A.}\ \bibnamefont
  {Mashkovich}}, \bibinfo {author} {\bibfnamefont {K.~A.}\ \bibnamefont
  {Grishunin}}, \bibinfo {author} {\bibfnamefont {R.~M.}\ \bibnamefont
  {Dubrovin}}, \bibinfo {author} {\bibfnamefont {A.~K.}\ \bibnamefont
  {Zvezdin}}, \bibinfo {author} {\bibfnamefont {R.~V.}\ \bibnamefont
  {Pisarev}},\ and\ \bibinfo {author} {\bibfnamefont {A.~V.}\ \bibnamefont
  {Kimel}},\ }\href {https://doi.org/10.1126/science.abk1121} {\bibfield
  {journal} {\bibinfo  {journal} {Science}\ }\textbf {\bibinfo {volume}
  {374}},\ \bibinfo {pages} {1608} (\bibinfo {year} {2021})}\BibitemShut
  {NoStop}%
\bibitem [{\citenamefont {Rongione}\ \emph {et~al.}(2023)\citenamefont
  {Rongione}, \citenamefont {Gueckstock}, \citenamefont {Mattern},
  \citenamefont {Gomonay}, \citenamefont {Meer}, \citenamefont {Schmitt},
  \citenamefont {Ramos}, \citenamefont {Kikkawa}, \citenamefont {Mi{\v{c}}ica},
  \citenamefont {Saitoh}, \citenamefont {Sinova}, \citenamefont {Jaffr{\`e}s},
  \citenamefont {Mangeney}, \citenamefont {Goennenwein}, \citenamefont
  {Gepr{\"a}gs}, \citenamefont {Kampfrath}, \citenamefont {Kl{\"a}ui},
  \citenamefont {Bargheer}, \citenamefont {Seifert}, \citenamefont {Dhillon},\
  and\ \citenamefont {Lebrun}}]{Rongione2023}%
  \BibitemOpen
  \bibfield  {author} {\bibinfo {author} {\bibfnamefont {E.}~\bibnamefont
  {Rongione}}, \bibinfo {author} {\bibfnamefont {O.}~\bibnamefont
  {Gueckstock}}, \bibinfo {author} {\bibfnamefont {M.}~\bibnamefont {Mattern}},
  \bibinfo {author} {\bibfnamefont {O.}~\bibnamefont {Gomonay}}, \bibinfo
  {author} {\bibfnamefont {H.}~\bibnamefont {Meer}}, \bibinfo {author}
  {\bibfnamefont {C.}~\bibnamefont {Schmitt}}, \bibinfo {author} {\bibfnamefont
  {R.}~\bibnamefont {Ramos}}, \bibinfo {author} {\bibfnamefont
  {T.}~\bibnamefont {Kikkawa}}, \bibinfo {author} {\bibfnamefont
  {M.}~\bibnamefont {Mi{\v{c}}ica}}, \bibinfo {author} {\bibfnamefont
  {E.}~\bibnamefont {Saitoh}}, \bibinfo {author} {\bibfnamefont
  {J.}~\bibnamefont {Sinova}}, \bibinfo {author} {\bibfnamefont
  {H.}~\bibnamefont {Jaffr{\`e}s}}, \bibinfo {author} {\bibfnamefont
  {J.}~\bibnamefont {Mangeney}}, \bibinfo {author} {\bibfnamefont {S.~T.~B.}\
  \bibnamefont {Goennenwein}}, \bibinfo {author} {\bibfnamefont
  {S.}~\bibnamefont {Gepr{\"a}gs}}, \bibinfo {author} {\bibfnamefont
  {T.}~\bibnamefont {Kampfrath}}, \bibinfo {author} {\bibfnamefont
  {M.}~\bibnamefont {Kl{\"a}ui}}, \bibinfo {author} {\bibfnamefont
  {M.}~\bibnamefont {Bargheer}}, \bibinfo {author} {\bibfnamefont {T.~S.}\
  \bibnamefont {Seifert}}, \bibinfo {author} {\bibfnamefont {S.}~\bibnamefont
  {Dhillon}},\ and\ \bibinfo {author} {\bibfnamefont {R.}~\bibnamefont
  {Lebrun}},\ }\href {https://doi.org/10.1038/s41467-023-37509-6} {\bibfield
  {journal} {\bibinfo  {journal} {Nature Communications}\ }\textbf {\bibinfo
  {volume} {14}},\ \bibinfo {pages} {1818} (\bibinfo {year}
  {2023})}\BibitemShut {NoStop}%
\bibitem [{\citenamefont {Park}\ \emph {et~al.}(2011)\citenamefont {Park},
  \citenamefont {Wunderlich}, \citenamefont {Mart{\'i}}, \citenamefont
  {Hol{\'y}}, \citenamefont {Kurosaki}, \citenamefont {Yamada}, \citenamefont
  {Yamamoto}, \citenamefont {Nishide}, \citenamefont {Hayakawa}, \citenamefont
  {Takahashi}, \citenamefont {Shick},\ and\ \citenamefont
  {Jungwirth}}]{Park2011}%
  \BibitemOpen
  \bibfield  {author} {\bibinfo {author} {\bibfnamefont {B.~G.}\ \bibnamefont
  {Park}}, \bibinfo {author} {\bibfnamefont {J.}~\bibnamefont {Wunderlich}},
  \bibinfo {author} {\bibfnamefont {X.}~\bibnamefont {Mart{\'i}}}, \bibinfo
  {author} {\bibfnamefont {V.}~\bibnamefont {Hol{\'y}}}, \bibinfo {author}
  {\bibfnamefont {Y.}~\bibnamefont {Kurosaki}}, \bibinfo {author}
  {\bibfnamefont {M.}~\bibnamefont {Yamada}}, \bibinfo {author} {\bibfnamefont
  {H.}~\bibnamefont {Yamamoto}}, \bibinfo {author} {\bibfnamefont
  {A.}~\bibnamefont {Nishide}}, \bibinfo {author} {\bibfnamefont
  {J.}~\bibnamefont {Hayakawa}}, \bibinfo {author} {\bibfnamefont
  {H.}~\bibnamefont {Takahashi}}, \bibinfo {author} {\bibfnamefont {A.~B.}\
  \bibnamefont {Shick}},\ and\ \bibinfo {author} {\bibfnamefont
  {T.}~\bibnamefont {Jungwirth}},\ }\href {https://doi.org/10.1038/nmat2983}
  {\bibfield  {journal} {\bibinfo  {journal} {Nature Materials}\ }\textbf
  {\bibinfo {volume} {10}},\ \bibinfo {pages} {347} (\bibinfo {year}
  {2011})}\BibitemShut {NoStop}%
\bibitem [{\citenamefont {Loth}\ \emph {et~al.}(2012)\citenamefont {Loth},
  \citenamefont {Baumann}, \citenamefont {Lutz}, \citenamefont {Eigler},\ and\
  \citenamefont {Heinrich}}]{Loth2012}%
  \BibitemOpen
  \bibfield  {author} {\bibinfo {author} {\bibfnamefont {S.}~\bibnamefont
  {Loth}}, \bibinfo {author} {\bibfnamefont {S.}~\bibnamefont {Baumann}},
  \bibinfo {author} {\bibfnamefont {C.~P.}\ \bibnamefont {Lutz}}, \bibinfo
  {author} {\bibfnamefont {D.~M.}\ \bibnamefont {Eigler}},\ and\ \bibinfo
  {author} {\bibfnamefont {A.~J.}\ \bibnamefont {Heinrich}},\ }\href
  {https://doi.org/10.1126/science.1214131} {\bibfield  {journal} {\bibinfo
  {journal} {Science}\ }\textbf {\bibinfo {volume} {335}},\ \bibinfo {pages}
  {196} (\bibinfo {year} {2012})}\BibitemShut {NoStop}%
\bibitem [{\citenamefont {Marti}\ \emph {et~al.}(2014)\citenamefont {Marti},
  \citenamefont {Fina}, \citenamefont {Frontera}, \citenamefont {Liu},
  \citenamefont {Wadley}, \citenamefont {He}, \citenamefont {Paull},
  \citenamefont {Clarkson}, \citenamefont {Kudrnovsk{\'y}}, \citenamefont
  {Turek}, \citenamefont {Kune{\v{s}}}, \citenamefont {Yi}, \citenamefont
  {Chu}, \citenamefont {Nelson}, \citenamefont {You}, \citenamefont {Arenholz},
  \citenamefont {Salahuddin}, \citenamefont {Fontcuberta}, \citenamefont
  {Jungwirth},\ and\ \citenamefont {Ramesh}}]{Marti2014}%
  \BibitemOpen
  \bibfield  {author} {\bibinfo {author} {\bibfnamefont {X.}~\bibnamefont
  {Marti}}, \bibinfo {author} {\bibfnamefont {I.}~\bibnamefont {Fina}},
  \bibinfo {author} {\bibfnamefont {C.}~\bibnamefont {Frontera}}, \bibinfo
  {author} {\bibfnamefont {J.}~\bibnamefont {Liu}}, \bibinfo {author}
  {\bibfnamefont {P.}~\bibnamefont {Wadley}}, \bibinfo {author} {\bibfnamefont
  {Q.}~\bibnamefont {He}}, \bibinfo {author} {\bibfnamefont {R.~J.}\
  \bibnamefont {Paull}}, \bibinfo {author} {\bibfnamefont {J.~D.}\ \bibnamefont
  {Clarkson}}, \bibinfo {author} {\bibfnamefont {J.}~\bibnamefont
  {Kudrnovsk{\'y}}}, \bibinfo {author} {\bibfnamefont {I.}~\bibnamefont
  {Turek}}, \bibinfo {author} {\bibfnamefont {J.}~\bibnamefont {Kune{\v{s}}}},
  \bibinfo {author} {\bibfnamefont {D.}~\bibnamefont {Yi}}, \bibinfo {author}
  {\bibfnamefont {J.-H.}\ \bibnamefont {Chu}}, \bibinfo {author} {\bibfnamefont
  {C.~T.}\ \bibnamefont {Nelson}}, \bibinfo {author} {\bibfnamefont
  {L.}~\bibnamefont {You}}, \bibinfo {author} {\bibfnamefont {E.}~\bibnamefont
  {Arenholz}}, \bibinfo {author} {\bibfnamefont {S.}~\bibnamefont
  {Salahuddin}}, \bibinfo {author} {\bibfnamefont {J.}~\bibnamefont
  {Fontcuberta}}, \bibinfo {author} {\bibfnamefont {T.}~\bibnamefont
  {Jungwirth}},\ and\ \bibinfo {author} {\bibfnamefont {R.}~\bibnamefont
  {Ramesh}},\ }\href {https://doi.org/10.1038/nmat3861} {\bibfield  {journal}
  {\bibinfo  {journal} {Nature Materials}\ }\textbf {\bibinfo {volume} {13}},\
  \bibinfo {pages} {367} (\bibinfo {year} {2014})}\BibitemShut {NoStop}%
\bibitem [{\citenamefont {Wadley}\ \emph {et~al.}(2016)\citenamefont {Wadley},
  \citenamefont {Howells}, \citenamefont {Železný}, \citenamefont {Andrews},
  \citenamefont {Hills}, \citenamefont {Campion}, \citenamefont {Novák},
  \citenamefont {Olejník}, \citenamefont {Maccherozzi}, \citenamefont {Dhesi},
  \citenamefont {Martin}, \citenamefont {Wagner}, \citenamefont {Wunderlich},
  \citenamefont {Freimuth}, \citenamefont {Mokrousov}, \citenamefont {Kuneš},
  \citenamefont {Chauhan}, \citenamefont {Grzybowski}, \citenamefont
  {Rushforth}, \citenamefont {Edmonds}, \citenamefont {Gallagher},\ and\
  \citenamefont {Jungwirth}}]{Wadley2016}%
  \BibitemOpen
  \bibfield  {author} {\bibinfo {author} {\bibfnamefont {P.}~\bibnamefont
  {Wadley}}, \bibinfo {author} {\bibfnamefont {B.}~\bibnamefont {Howells}},
  \bibinfo {author} {\bibfnamefont {J.}~\bibnamefont {Železný}}, \bibinfo
  {author} {\bibfnamefont {C.}~\bibnamefont {Andrews}}, \bibinfo {author}
  {\bibfnamefont {V.}~\bibnamefont {Hills}}, \bibinfo {author} {\bibfnamefont
  {R.~P.}\ \bibnamefont {Campion}}, \bibinfo {author} {\bibfnamefont
  {V.}~\bibnamefont {Novák}}, \bibinfo {author} {\bibfnamefont
  {K.}~\bibnamefont {Olejník}}, \bibinfo {author} {\bibfnamefont
  {F.}~\bibnamefont {Maccherozzi}}, \bibinfo {author} {\bibfnamefont {S.~S.}\
  \bibnamefont {Dhesi}}, \bibinfo {author} {\bibfnamefont {S.~Y.}\ \bibnamefont
  {Martin}}, \bibinfo {author} {\bibfnamefont {T.}~\bibnamefont {Wagner}},
  \bibinfo {author} {\bibfnamefont {J.}~\bibnamefont {Wunderlich}}, \bibinfo
  {author} {\bibfnamefont {F.}~\bibnamefont {Freimuth}}, \bibinfo {author}
  {\bibfnamefont {Y.}~\bibnamefont {Mokrousov}}, \bibinfo {author}
  {\bibfnamefont {J.}~\bibnamefont {Kuneš}}, \bibinfo {author} {\bibfnamefont
  {J.~S.}\ \bibnamefont {Chauhan}}, \bibinfo {author} {\bibfnamefont {M.~J.}\
  \bibnamefont {Grzybowski}}, \bibinfo {author} {\bibfnamefont {A.~W.}\
  \bibnamefont {Rushforth}}, \bibinfo {author} {\bibfnamefont {K.~W.}\
  \bibnamefont {Edmonds}}, \bibinfo {author} {\bibfnamefont {B.~L.}\
  \bibnamefont {Gallagher}},\ and\ \bibinfo {author} {\bibfnamefont
  {T.}~\bibnamefont {Jungwirth}},\ }\href
  {https://doi.org/10.1126/science.aab1031} {\bibfield  {journal} {\bibinfo
  {journal} {Science}\ }\textbf {\bibinfo {volume} {351}},\ \bibinfo {pages}
  {587} (\bibinfo {year} {2016})}\BibitemShut {NoStop}%
\bibitem [{\citenamefont {Kosub}\ \emph {et~al.}(2017)\citenamefont {Kosub},
  \citenamefont {Kopte}, \citenamefont {Hühne}, \citenamefont {Appel},
  \citenamefont {Shields}, \citenamefont {Maletinsky}, \citenamefont {Hübner},
  \citenamefont {Liedke}, \citenamefont {Fassbender}, \citenamefont {Schmidt},\
  and\ \citenamefont {Makarov}}]{Kosub2017}%
  \BibitemOpen
  \bibfield  {author} {\bibinfo {author} {\bibfnamefont {T.}~\bibnamefont
  {Kosub}}, \bibinfo {author} {\bibfnamefont {M.}~\bibnamefont {Kopte}},
  \bibinfo {author} {\bibfnamefont {R.}~\bibnamefont {Hühne}}, \bibinfo
  {author} {\bibfnamefont {P.}~\bibnamefont {Appel}}, \bibinfo {author}
  {\bibfnamefont {B.}~\bibnamefont {Shields}}, \bibinfo {author} {\bibfnamefont
  {P.}~\bibnamefont {Maletinsky}}, \bibinfo {author} {\bibfnamefont
  {R.}~\bibnamefont {Hübner}}, \bibinfo {author} {\bibfnamefont
  {M.}~\bibnamefont {Liedke}}, \bibinfo {author} {\bibfnamefont
  {J.}~\bibnamefont {Fassbender}}, \bibinfo {author} {\bibfnamefont
  {O.}~\bibnamefont {Schmidt}},\ and\ \bibinfo {author} {\bibfnamefont
  {D.}~\bibnamefont {Makarov}},\ }\href
  {https://doi.org/https://doi.org/10.1038/ncomms13985} {\bibfield  {journal}
  {\bibinfo  {journal} {Nature Communications}\ }\textbf {\bibinfo {volume}
  {8}},\ \bibinfo {pages} {13985} (\bibinfo {year} {2017})}\BibitemShut
  {NoStop}%
\bibitem [{\citenamefont {N\'u\~nez}\ \emph {et~al.}(2006)\citenamefont
  {N\'u\~nez}, \citenamefont {Duine}, \citenamefont {Haney},\ and\
  \citenamefont {MacDonald}}]{Nunez2006}%
  \BibitemOpen
  \bibfield  {author} {\bibinfo {author} {\bibfnamefont {A.~S.}\ \bibnamefont
  {N\'u\~nez}}, \bibinfo {author} {\bibfnamefont {R.~A.}\ \bibnamefont
  {Duine}}, \bibinfo {author} {\bibfnamefont {P.}~\bibnamefont {Haney}},\ and\
  \bibinfo {author} {\bibfnamefont {A.~H.}\ \bibnamefont {MacDonald}},\ }\href
  {https://doi.org/10.1103/PhysRevB.73.214426} {\bibfield  {journal} {\bibinfo
  {journal} {Phys. Rev. B}\ }\textbf {\bibinfo {volume} {73}},\ \bibinfo
  {pages} {214426} (\bibinfo {year} {2006})}\BibitemShut {NoStop}%
\bibitem [{\citenamefont {Merodio}\ \emph {et~al.}(2014)\citenamefont
  {Merodio}, \citenamefont {Kalitsov}, \citenamefont {Béa}, \citenamefont
  {Baltz},\ and\ \citenamefont {Chshiev}}]{Merodio2014}%
  \BibitemOpen
  \bibfield  {author} {\bibinfo {author} {\bibfnamefont {P.}~\bibnamefont
  {Merodio}}, \bibinfo {author} {\bibfnamefont {A.}~\bibnamefont {Kalitsov}},
  \bibinfo {author} {\bibfnamefont {H.}~\bibnamefont {Béa}}, \bibinfo {author}
  {\bibfnamefont {V.}~\bibnamefont {Baltz}},\ and\ \bibinfo {author}
  {\bibfnamefont {M.}~\bibnamefont {Chshiev}},\ }\href
  {https://doi.org/10.1063/1.4896291} {\bibfield  {journal} {\bibinfo
  {journal} {Applied Physics Letters}\ }\textbf {\bibinfo {volume} {105}},\
  \bibinfo {pages} {122403} (\bibinfo {year} {2014})}\BibitemShut {NoStop}%
\bibitem [{\citenamefont {Lin}\ \emph {et~al.}(2019)\citenamefont {Lin},
  \citenamefont {Yang}, \citenamefont {Tsai}, \citenamefont {Chen},
  \citenamefont {Huang}, \citenamefont {Lin},\ and\ \citenamefont
  {Lai}}]{Lin2019}%
  \BibitemOpen
  \bibfield  {author} {\bibinfo {author} {\bibfnamefont {P.-H.}\ \bibnamefont
  {Lin}}, \bibinfo {author} {\bibfnamefont {B.-Y.}\ \bibnamefont {Yang}},
  \bibinfo {author} {\bibfnamefont {M.-H.}\ \bibnamefont {Tsai}}, \bibinfo
  {author} {\bibfnamefont {P.-C.}\ \bibnamefont {Chen}}, \bibinfo {author}
  {\bibfnamefont {K.-F.}\ \bibnamefont {Huang}}, \bibinfo {author}
  {\bibfnamefont {H.-H.}\ \bibnamefont {Lin}},\ and\ \bibinfo {author}
  {\bibfnamefont {C.-H.}\ \bibnamefont {Lai}},\ }\href
  {https://doi.org/10.1038/s41563-019-0289-4} {\bibfield  {journal} {\bibinfo
  {journal} {Nature Materials}\ }\textbf {\bibinfo {volume} {18}},\ \bibinfo
  {pages} {335} (\bibinfo {year} {2019})}\BibitemShut {NoStop}%
\bibitem [{\citenamefont {Liu}\ \emph {et~al.}(2020)\citenamefont {Liu},
  \citenamefont {Edmonds}, \citenamefont {Zhou},\ and\ \citenamefont
  {Wang}}]{Liu2020}%
  \BibitemOpen
  \bibfield  {author} {\bibinfo {author} {\bibfnamefont {X.~H.}\ \bibnamefont
  {Liu}}, \bibinfo {author} {\bibfnamefont {K.~W.}\ \bibnamefont {Edmonds}},
  \bibinfo {author} {\bibfnamefont {Z.~P.}\ \bibnamefont {Zhou}},\ and\
  \bibinfo {author} {\bibfnamefont {K.~Y.}\ \bibnamefont {Wang}},\ }\href
  {https://doi.org/10.1103/PhysRevApplied.13.014059} {\bibfield  {journal}
  {\bibinfo  {journal} {Phys. Rev. Appl.}\ }\textbf {\bibinfo {volume} {13}},\
  \bibinfo {pages} {014059} (\bibinfo {year} {2020})}\BibitemShut {NoStop}%
\bibitem [{\citenamefont {Zhang}\ \emph {et~al.}(2021)\citenamefont {Zhang},
  \citenamefont {Deng}, \citenamefont {Liu}, \citenamefont {Zhan},
  \citenamefont {Zhu},\ and\ \citenamefont {Wang}}]{Zhang2021}%
  \BibitemOpen
  \bibfield  {author} {\bibinfo {author} {\bibfnamefont {E.~Z.}\ \bibnamefont
  {Zhang}}, \bibinfo {author} {\bibfnamefont {Y.~C.}\ \bibnamefont {Deng}},
  \bibinfo {author} {\bibfnamefont {X.~H.}\ \bibnamefont {Liu}}, \bibinfo
  {author} {\bibfnamefont {X.~Z.}\ \bibnamefont {Zhan}}, \bibinfo {author}
  {\bibfnamefont {T.}~\bibnamefont {Zhu}},\ and\ \bibinfo {author}
  {\bibfnamefont {K.~Y.}\ \bibnamefont {Wang}},\ }\href
  {https://doi.org/10.1103/PhysRevB.104.134408} {\bibfield  {journal} {\bibinfo
   {journal} {Phys. Rev. B}\ }\textbf {\bibinfo {volume} {104}},\ \bibinfo
  {pages} {134408} (\bibinfo {year} {2021})}\BibitemShut {NoStop}%
\bibitem [{\citenamefont {Roy}\ \emph {et~al.}(2005)\citenamefont {Roy},
  \citenamefont {Fitzsimmons}, \citenamefont {Park}, \citenamefont {Dorn},
  \citenamefont {Petracic}, \citenamefont {Roshchin}, \citenamefont {Li},
  \citenamefont {Batlle}, \citenamefont {Morales}, \citenamefont {Misra},
  \citenamefont {Zhang}, \citenamefont {Chesnel}, \citenamefont {Kortright},
  \citenamefont {Sinha},\ and\ \citenamefont {Schuller}}]{Roy2005}%
  \BibitemOpen
  \bibfield  {author} {\bibinfo {author} {\bibfnamefont {S.}~\bibnamefont
  {Roy}}, \bibinfo {author} {\bibfnamefont {M.~R.}\ \bibnamefont
  {Fitzsimmons}}, \bibinfo {author} {\bibfnamefont {S.}~\bibnamefont {Park}},
  \bibinfo {author} {\bibfnamefont {M.}~\bibnamefont {Dorn}}, \bibinfo {author}
  {\bibfnamefont {O.}~\bibnamefont {Petracic}}, \bibinfo {author}
  {\bibfnamefont {I.~V.}\ \bibnamefont {Roshchin}}, \bibinfo {author}
  {\bibfnamefont {Z.-P.}\ \bibnamefont {Li}}, \bibinfo {author} {\bibfnamefont
  {X.}~\bibnamefont {Batlle}}, \bibinfo {author} {\bibfnamefont
  {R.}~\bibnamefont {Morales}}, \bibinfo {author} {\bibfnamefont
  {A.}~\bibnamefont {Misra}}, \bibinfo {author} {\bibfnamefont
  {X.}~\bibnamefont {Zhang}}, \bibinfo {author} {\bibfnamefont
  {K.}~\bibnamefont {Chesnel}}, \bibinfo {author} {\bibfnamefont {J.~B.}\
  \bibnamefont {Kortright}}, \bibinfo {author} {\bibfnamefont {S.~K.}\
  \bibnamefont {Sinha}},\ and\ \bibinfo {author} {\bibfnamefont {I.~K.}\
  \bibnamefont {Schuller}},\ }\href
  {https://doi.org/10.1103/PhysRevLett.95.047201} {\bibfield  {journal}
  {\bibinfo  {journal} {Phys. Rev. Lett.}\ }\textbf {\bibinfo {volume} {95}},\
  \bibinfo {pages} {047201} (\bibinfo {year} {2005})}\BibitemShut {NoStop}%
\bibitem [{\citenamefont {Zhou}\ \emph {et~al.}(2015)\citenamefont {Zhou},
  \citenamefont {Ma}, \citenamefont {Shi}, \citenamefont {Fan}, \citenamefont
  {Evans}, \citenamefont {Zheng}, \citenamefont {Chantrell}, \citenamefont
  {Mangin}, \citenamefont {Zhang},\ and\ \citenamefont {Zhou}}]{Zhou2015}%
  \BibitemOpen
  \bibfield  {author} {\bibinfo {author} {\bibfnamefont {X.}~\bibnamefont
  {Zhou}}, \bibinfo {author} {\bibfnamefont {L.}~\bibnamefont {Ma}}, \bibinfo
  {author} {\bibfnamefont {Z.}~\bibnamefont {Shi}}, \bibinfo {author}
  {\bibfnamefont {W.~J.}\ \bibnamefont {Fan}}, \bibinfo {author} {\bibfnamefont
  {R.~F.~L.}\ \bibnamefont {Evans}}, \bibinfo {author} {\bibfnamefont {J.-G.}\
  \bibnamefont {Zheng}}, \bibinfo {author} {\bibfnamefont {R.~W.}\ \bibnamefont
  {Chantrell}}, \bibinfo {author} {\bibfnamefont {S.}~\bibnamefont {Mangin}},
  \bibinfo {author} {\bibfnamefont {H.~W.}\ \bibnamefont {Zhang}},\ and\
  \bibinfo {author} {\bibfnamefont {S.~M.}\ \bibnamefont {Zhou}},\ }\href
  {https://doi.org/10.1038/srep09183} {\bibfield  {journal} {\bibinfo
  {journal} {Scientific Reports}\ }\textbf {\bibinfo {volume} {5}},\ \bibinfo
  {pages} {9183} (\bibinfo {year} {2015})}\BibitemShut {NoStop}%
\bibitem [{\citenamefont {Lapa}\ \emph {et~al.}(2020)\citenamefont {Lapa},
  \citenamefont {Lee}, \citenamefont {Roshchin}, \citenamefont {Belashchenko},\
  and\ \citenamefont {Schuller}}]{Lapa2020}%
  \BibitemOpen
  \bibfield  {author} {\bibinfo {author} {\bibfnamefont {P.~N.}\ \bibnamefont
  {Lapa}}, \bibinfo {author} {\bibfnamefont {M.-H.}\ \bibnamefont {Lee}},
  \bibinfo {author} {\bibfnamefont {I.~V.}\ \bibnamefont {Roshchin}}, \bibinfo
  {author} {\bibfnamefont {K.~D.}\ \bibnamefont {Belashchenko}},\ and\ \bibinfo
  {author} {\bibfnamefont {I.~K.}\ \bibnamefont {Schuller}},\ }\href
  {https://doi.org/10.1103/PhysRevB.102.174406} {\bibfield  {journal} {\bibinfo
   {journal} {Phys. Rev. B}\ }\textbf {\bibinfo {volume} {102}},\ \bibinfo
  {pages} {174406} (\bibinfo {year} {2020})}\BibitemShut {NoStop}%
\bibitem [{\citenamefont {Gentile}\ \emph {et~al.}(2022)\citenamefont
  {Gentile}, \citenamefont {Cuoco}, \citenamefont {Volkov}, \citenamefont
  {Ying}, \citenamefont {Vera-Marun}, \citenamefont {Makarov},\ and\
  \citenamefont {Ortix}}]{Gentile2022}%
  \BibitemOpen
  \bibfield  {author} {\bibinfo {author} {\bibfnamefont {P.}~\bibnamefont
  {Gentile}}, \bibinfo {author} {\bibfnamefont {M.}~\bibnamefont {Cuoco}},
  \bibinfo {author} {\bibfnamefont {O.~M.}\ \bibnamefont {Volkov}}, \bibinfo
  {author} {\bibfnamefont {Z.-J.}\ \bibnamefont {Ying}}, \bibinfo {author}
  {\bibfnamefont {I.~J.}\ \bibnamefont {Vera-Marun}}, \bibinfo {author}
  {\bibfnamefont {D.}~\bibnamefont {Makarov}},\ and\ \bibinfo {author}
  {\bibfnamefont {C.}~\bibnamefont {Ortix}},\ }\href
  {https://doi.org/10.1038/s41928-022-00820-z} {\bibfield  {journal} {\bibinfo
  {journal} {Nature Electronics}\ }\textbf {\bibinfo {volume} {5}},\ \bibinfo
  {pages} {551} (\bibinfo {year} {2022})}\BibitemShut {NoStop}%
\bibitem [{\citenamefont {Streubel}\ \emph {et~al.}(2016)\citenamefont
  {Streubel}, \citenamefont {Fischer}, \citenamefont {Kronast}, \citenamefont
  {Kravchuk}, \citenamefont {Sheka}, \citenamefont {Gaididei}, \citenamefont
  {Schmidt},\ and\ \citenamefont {Makarov}}]{Streubel2016}%
  \BibitemOpen
  \bibfield  {author} {\bibinfo {author} {\bibfnamefont {R.}~\bibnamefont
  {Streubel}}, \bibinfo {author} {\bibfnamefont {P.}~\bibnamefont {Fischer}},
  \bibinfo {author} {\bibfnamefont {F.}~\bibnamefont {Kronast}}, \bibinfo
  {author} {\bibfnamefont {V.~P.}\ \bibnamefont {Kravchuk}}, \bibinfo {author}
  {\bibfnamefont {D.~D.}\ \bibnamefont {Sheka}}, \bibinfo {author}
  {\bibfnamefont {Y.}~\bibnamefont {Gaididei}}, \bibinfo {author}
  {\bibfnamefont {O.~G.}\ \bibnamefont {Schmidt}},\ and\ \bibinfo {author}
  {\bibfnamefont {D.}~\bibnamefont {Makarov}},\ }\href
  {https://doi.org/10.1088/0022-3727/49/36/363001} {\bibfield  {journal}
  {\bibinfo  {journal} {Journal of Physics D: Applied Physics}\ }\textbf
  {\bibinfo {volume} {49}},\ \bibinfo {pages} {363001} (\bibinfo {year}
  {2016})}\BibitemShut {NoStop}%
\bibitem [{\citenamefont {Streubel}\ \emph {et~al.}(2021)\citenamefont
  {Streubel}, \citenamefont {Tsymbal},\ and\ \citenamefont
  {Fischer}}]{Streubel2021}%
  \BibitemOpen
  \bibfield  {author} {\bibinfo {author} {\bibfnamefont {R.}~\bibnamefont
  {Streubel}}, \bibinfo {author} {\bibfnamefont {E.~Y.}\ \bibnamefont
  {Tsymbal}},\ and\ \bibinfo {author} {\bibfnamefont {P.}~\bibnamefont
  {Fischer}},\ }\href {https://doi.org/https://doi.org/10.1063/5.0054025}
  {\bibfield  {journal} {\bibinfo  {journal} {Journal of Applied Physics}\
  }\textbf {\bibinfo {volume} {129}},\ \bibinfo {pages} {210902} (\bibinfo
  {year} {2021})}\BibitemShut {NoStop}%
\bibitem [{\citenamefont {Gentile}\ \emph {et~al.}(2013)\citenamefont
  {Gentile}, \citenamefont {Cuoco},\ and\ \citenamefont {Ortix}}]{Gentile2013}%
  \BibitemOpen
  \bibfield  {author} {\bibinfo {author} {\bibfnamefont {P.}~\bibnamefont
  {Gentile}}, \bibinfo {author} {\bibfnamefont {M.}~\bibnamefont {Cuoco}},\
  and\ \bibinfo {author} {\bibfnamefont {C.}~\bibnamefont {Ortix}},\ }\href
  {https://doi.org/10.1142/S201032471340002X} {\bibfield  {journal} {\bibinfo
  {journal} {Spin}\ }\textbf {\bibinfo {volume} {3}},\ \bibinfo {pages}
  {1340002} (\bibinfo {year} {2013})}\BibitemShut {NoStop}%
\bibitem [{\citenamefont {Gentile}\ \emph {et~al.}(2015)\citenamefont
  {Gentile}, \citenamefont {Cuoco},\ and\ \citenamefont {Ortix}}]{Gentile2015}%
  \BibitemOpen
  \bibfield  {author} {\bibinfo {author} {\bibfnamefont {P.}~\bibnamefont
  {Gentile}}, \bibinfo {author} {\bibfnamefont {M.}~\bibnamefont {Cuoco}},\
  and\ \bibinfo {author} {\bibfnamefont {C.}~\bibnamefont {Ortix}},\ }\href
  {https://doi.org/10.1103/PhysRevLett.115.256801} {\bibfield  {journal}
  {\bibinfo  {journal} {Phys. Rev. Lett.}\ }\textbf {\bibinfo {volume} {115}},\
  \bibinfo {pages} {256801} (\bibinfo {year} {2015})}\BibitemShut {NoStop}%
\bibitem [{\citenamefont {Ortix}\ and\ \citenamefont {van~den
  Brink}(2010)}]{Ortix2010}%
  \BibitemOpen
  \bibfield  {author} {\bibinfo {author} {\bibfnamefont {C.}~\bibnamefont
  {Ortix}}\ and\ \bibinfo {author} {\bibfnamefont {J.}~\bibnamefont {van~den
  Brink}},\ }\href {https://doi.org/10.1103/PhysRevB.81.165419} {\bibfield
  {journal} {\bibinfo  {journal} {Phys. Rev. B}\ }\textbf {\bibinfo {volume}
  {81}},\ \bibinfo {pages} {165419} (\bibinfo {year} {2010})}\BibitemShut
  {NoStop}%
\bibitem [{\citenamefont {Ortix}\ \emph {et~al.}(2011)\citenamefont {Ortix},
  \citenamefont {Kiravittaya}, \citenamefont {Schmidt},\ and\ \citenamefont
  {van~den Brink}}]{Ortix2011}%
  \BibitemOpen
  \bibfield  {author} {\bibinfo {author} {\bibfnamefont {C.}~\bibnamefont
  {Ortix}}, \bibinfo {author} {\bibfnamefont {S.}~\bibnamefont {Kiravittaya}},
  \bibinfo {author} {\bibfnamefont {O.~G.}\ \bibnamefont {Schmidt}},\ and\
  \bibinfo {author} {\bibfnamefont {J.}~\bibnamefont {van~den Brink}},\ }\href
  {https://doi.org/10.1103/PhysRevB.84.045438} {\bibfield  {journal} {\bibinfo
  {journal} {Physical Review B}\ }\textbf {\bibinfo {volume} {84}},\ \bibinfo
  {pages} {045438} (\bibinfo {year} {2011})}\BibitemShut {NoStop}%
\bibitem [{\citenamefont {Francica}\ \emph {et~al.}(2020)\citenamefont
  {Francica}, \citenamefont {Cuoco},\ and\ \citenamefont
  {Gentile}}]{Francica2020}%
  \BibitemOpen
  \bibfield  {author} {\bibinfo {author} {\bibfnamefont {G.}~\bibnamefont
  {Francica}}, \bibinfo {author} {\bibfnamefont {M.}~\bibnamefont {Cuoco}},\
  and\ \bibinfo {author} {\bibfnamefont {P.}~\bibnamefont {Gentile}},\ }\href
  {https://doi.org/10.1103/PhysRevB.101.094504} {\bibfield  {journal} {\bibinfo
   {journal} {Phys. Rev. B}\ }\textbf {\bibinfo {volume} {101}},\ \bibinfo
  {pages} {094504} (\bibinfo {year} {2020})}\BibitemShut {NoStop}%
\bibitem [{\citenamefont {Volkov}\ \emph {et~al.}(2019)\citenamefont {Volkov},
  \citenamefont {K\'akay}, \citenamefont {Kronast}, \citenamefont {M\"onch},
  \citenamefont {Mawass}, \citenamefont {Fassbender},\ and\ \citenamefont
  {Makarov}}]{Volkov2019}%
  \BibitemOpen
  \bibfield  {author} {\bibinfo {author} {\bibfnamefont {O.~M.}\ \bibnamefont
  {Volkov}}, \bibinfo {author} {\bibfnamefont {A.}~\bibnamefont {K\'akay}},
  \bibinfo {author} {\bibfnamefont {F.}~\bibnamefont {Kronast}}, \bibinfo
  {author} {\bibfnamefont {I.}~\bibnamefont {M\"onch}}, \bibinfo {author}
  {\bibfnamefont {M.-A.}\ \bibnamefont {Mawass}}, \bibinfo {author}
  {\bibfnamefont {J.}~\bibnamefont {Fassbender}},\ and\ \bibinfo {author}
  {\bibfnamefont {D.}~\bibnamefont {Makarov}},\ }\href
  {https://doi.org/10.1103/PhysRevLett.123.077201} {\bibfield  {journal}
  {\bibinfo  {journal} {Phys. Rev. Lett.}\ }\textbf {\bibinfo {volume} {123}},\
  \bibinfo {pages} {077201} (\bibinfo {year} {2019})}\BibitemShut {NoStop}%
\bibitem [{\citenamefont {Frustaglia}\ and\ \citenamefont
  {Nitta}(2020)}]{Frustaglia2020}%
  \BibitemOpen
  \bibfield  {author} {\bibinfo {author} {\bibfnamefont {D.}~\bibnamefont
  {Frustaglia}}\ and\ \bibinfo {author} {\bibfnamefont {J.}~\bibnamefont
  {Nitta}},\ }\href {https://doi.org/https://doi.org/10.1016/j.ssc.2020.113864}
  {\bibfield  {journal} {\bibinfo  {journal} {Solid State Communications}\
  }\textbf {\bibinfo {volume} {311}},\ \bibinfo {pages} {113864} (\bibinfo
  {year} {2020})}\BibitemShut {NoStop}%
\bibitem [{\citenamefont {Rodr\'iguez}\ and\ \citenamefont
  {Frustaglia}(2021)}]{Rodriguez2021}%
  \BibitemOpen
  \bibfield  {author} {\bibinfo {author} {\bibfnamefont {E.~J.}\ \bibnamefont
  {Rodr\'iguez}}\ and\ \bibinfo {author} {\bibfnamefont {D.}~\bibnamefont
  {Frustaglia}},\ }\href {https://doi.org/10.1103/PhysRevB.104.195308}
  {\bibfield  {journal} {\bibinfo  {journal} {Phys. Rev. B}\ }\textbf {\bibinfo
  {volume} {104}},\ \bibinfo {pages} {195308} (\bibinfo {year}
  {2021})}\BibitemShut {NoStop}%
\bibitem [{\citenamefont {Kutlin}\ and\ \citenamefont
  {Mel'nikov}(2020)}]{Kutlin2020}%
  \BibitemOpen
  \bibfield  {author} {\bibinfo {author} {\bibfnamefont {A.~G.}\ \bibnamefont
  {Kutlin}}\ and\ \bibinfo {author} {\bibfnamefont {A.~S.}\ \bibnamefont
  {Mel'nikov}},\ }\href {https://doi.org/10.1103/PhysRevB.101.045418}
  {\bibfield  {journal} {\bibinfo  {journal} {Phys. Rev. B}\ }\textbf {\bibinfo
  {volume} {101}},\ \bibinfo {pages} {045418} (\bibinfo {year}
  {2020})}\BibitemShut {NoStop}%
\bibitem [{\citenamefont {Chou}\ \emph {et~al.}(2021)\citenamefont {Chou},
  \citenamefont {Chen}, \citenamefont {Liu}, \citenamefont {Chung},\ and\
  \citenamefont {Mou}}]{Chou2021}%
  \BibitemOpen
  \bibfield  {author} {\bibinfo {author} {\bibfnamefont {P.-H.}\ \bibnamefont
  {Chou}}, \bibinfo {author} {\bibfnamefont {C.-H.}\ \bibnamefont {Chen}},
  \bibinfo {author} {\bibfnamefont {S.-W.}\ \bibnamefont {Liu}}, \bibinfo
  {author} {\bibfnamefont {C.-H.}\ \bibnamefont {Chung}},\ and\ \bibinfo
  {author} {\bibfnamefont {C.-Y.}\ \bibnamefont {Mou}},\ }\href
  {https://doi.org/10.1103/PhysRevB.103.014508} {\bibfield  {journal} {\bibinfo
   {journal} {Phys. Rev. B}\ }\textbf {\bibinfo {volume} {103}},\ \bibinfo
  {pages} {014508} (\bibinfo {year} {2021})}\BibitemShut {NoStop}%
\bibitem [{\citenamefont {Wang}\ \emph {et~al.}(2019)\citenamefont {Wang},
  \citenamefont {Saarikoski}, \citenamefont {Reynoso}, \citenamefont
  {Baltan\'as}, \citenamefont {Frustaglia},\ and\ \citenamefont
  {Nitta}}]{Wang2019}%
  \BibitemOpen
  \bibfield  {author} {\bibinfo {author} {\bibfnamefont {M.}~\bibnamefont
  {Wang}}, \bibinfo {author} {\bibfnamefont {H.}~\bibnamefont {Saarikoski}},
  \bibinfo {author} {\bibfnamefont {A.~A.}\ \bibnamefont {Reynoso}}, \bibinfo
  {author} {\bibfnamefont {J.~P.}\ \bibnamefont {Baltan\'as}}, \bibinfo
  {author} {\bibfnamefont {D.}~\bibnamefont {Frustaglia}},\ and\ \bibinfo
  {author} {\bibfnamefont {J.}~\bibnamefont {Nitta}},\ }\href
  {https://doi.org/10.1103/PhysRevLett.123.266804} {\bibfield  {journal}
  {\bibinfo  {journal} {Phys. Rev. Lett.}\ }\textbf {\bibinfo {volume} {123}},\
  \bibinfo {pages} {266804} (\bibinfo {year} {2019})}\BibitemShut {NoStop}%
\bibitem [{\citenamefont {Nagasawa}\ \emph {et~al.}(2013)\citenamefont
  {Nagasawa}, \citenamefont {Frustaglia}, \citenamefont {Saarikoski},
  \citenamefont {Richter},\ and\ \citenamefont {Nitta}}]{Nagasawa2013}%
  \BibitemOpen
  \bibfield  {author} {\bibinfo {author} {\bibfnamefont {F.}~\bibnamefont
  {Nagasawa}}, \bibinfo {author} {\bibfnamefont {D.}~\bibnamefont
  {Frustaglia}}, \bibinfo {author} {\bibfnamefont {H.}~\bibnamefont
  {Saarikoski}}, \bibinfo {author} {\bibfnamefont {K.}~\bibnamefont
  {Richter}},\ and\ \bibinfo {author} {\bibfnamefont {J.}~\bibnamefont
  {Nitta}},\ }\href {https://doi.org/10.1038/ncomms3526} {\bibfield  {journal}
  {\bibinfo  {journal} {Nature communications}\ }\textbf {\bibinfo {volume}
  {4}},\ \bibinfo {pages} {1} (\bibinfo {year} {2013})}\BibitemShut {NoStop}%
\bibitem [{\citenamefont {Ying}\ \emph {et~al.}(2020)\citenamefont {Ying},
  \citenamefont {Gentile}, \citenamefont {Baltan\'as}, \citenamefont
  {Frustaglia}, \citenamefont {Ortix},\ and\ \citenamefont {Cuoco}}]{Ying2020}%
  \BibitemOpen
  \bibfield  {author} {\bibinfo {author} {\bibfnamefont {Z.-J.}\ \bibnamefont
  {Ying}}, \bibinfo {author} {\bibfnamefont {P.}~\bibnamefont {Gentile}},
  \bibinfo {author} {\bibfnamefont {J.~P.}\ \bibnamefont {Baltan\'as}},
  \bibinfo {author} {\bibfnamefont {D.}~\bibnamefont {Frustaglia}}, \bibinfo
  {author} {\bibfnamefont {C.}~\bibnamefont {Ortix}},\ and\ \bibinfo {author}
  {\bibfnamefont {M.}~\bibnamefont {Cuoco}},\ }\href
  {https://doi.org/10.1103/PhysRevResearch.2.023167} {\bibfield  {journal}
  {\bibinfo  {journal} {Phys. Rev. Res.}\ }\textbf {\bibinfo {volume} {2}},\
  \bibinfo {pages} {023167} (\bibinfo {year} {2020})}\BibitemShut {NoStop}%
\bibitem [{\citenamefont {Das}\ \emph {et~al.}(2019)\citenamefont {Das},
  \citenamefont {Makarov}, \citenamefont {Gentile}, \citenamefont {Cuoco},
  \citenamefont {van Wees}, \citenamefont {Ortix},\ and\ \citenamefont
  {Vera-Marun}}]{Das2019}%
  \BibitemOpen
  \bibfield  {author} {\bibinfo {author} {\bibfnamefont {K.~S.}\ \bibnamefont
  {Das}}, \bibinfo {author} {\bibfnamefont {D.}~\bibnamefont {Makarov}},
  \bibinfo {author} {\bibfnamefont {P.}~\bibnamefont {Gentile}}, \bibinfo
  {author} {\bibfnamefont {M.}~\bibnamefont {Cuoco}}, \bibinfo {author}
  {\bibfnamefont {B.~J.}\ \bibnamefont {van Wees}}, \bibinfo {author}
  {\bibfnamefont {C.}~\bibnamefont {Ortix}},\ and\ \bibinfo {author}
  {\bibfnamefont {I.~J.}\ \bibnamefont {Vera-Marun}},\ }\href
  {https://doi.org/10.1021/acs.nanolett.9b01994} {\bibfield  {journal}
  {\bibinfo  {journal} {Nano Letters}\ }\textbf {\bibinfo {volume} {19}},\
  \bibinfo {pages} {6839} (\bibinfo {year} {2019})}\BibitemShut {NoStop}%
\bibitem [{\citenamefont {Salamone}\ \emph {et~al.}(2021)\citenamefont
  {Salamone}, \citenamefont {Svendsen}, \citenamefont {Amundsen},\ and\
  \citenamefont {Jacobsen}}]{Salamone2021}%
  \BibitemOpen
  \bibfield  {author} {\bibinfo {author} {\bibfnamefont {T.}~\bibnamefont
  {Salamone}}, \bibinfo {author} {\bibfnamefont {M.~B.~M.}\ \bibnamefont
  {Svendsen}}, \bibinfo {author} {\bibfnamefont {M.}~\bibnamefont {Amundsen}},\
  and\ \bibinfo {author} {\bibfnamefont {S.}~\bibnamefont {Jacobsen}},\ }\href
  {https://doi.org/10.1103/PhysRevB.104.L060505} {\bibfield  {journal}
  {\bibinfo  {journal} {Phys. Rev. B}\ }\textbf {\bibinfo {volume} {104}},\
  \bibinfo {pages} {L060505} (\bibinfo {year} {2021})}\BibitemShut {NoStop}%
\bibitem [{\citenamefont {Salamone}\ \emph {et~al.}(2022)\citenamefont
  {Salamone}, \citenamefont {Hugdal}, \citenamefont {Amundsen},\ and\
  \citenamefont {Jacobsen}}]{Salamone2022}%
  \BibitemOpen
  \bibfield  {author} {\bibinfo {author} {\bibfnamefont {T.}~\bibnamefont
  {Salamone}}, \bibinfo {author} {\bibfnamefont {H.~G.}\ \bibnamefont
  {Hugdal}}, \bibinfo {author} {\bibfnamefont {M.}~\bibnamefont {Amundsen}},\
  and\ \bibinfo {author} {\bibfnamefont {S.~H.}\ \bibnamefont {Jacobsen}},\
  }\href {https://doi.org/10.1103/PhysRevB.105.134511} {\bibfield  {journal}
  {\bibinfo  {journal} {Phys. Rev. B}\ }\textbf {\bibinfo {volume} {105}},\
  \bibinfo {pages} {134511} (\bibinfo {year} {2022})}\BibitemShut {NoStop}%
\bibitem [{\citenamefont {Kundys}(2015)}]{Kundys2015}%
  \BibitemOpen
  \bibfield  {author} {\bibinfo {author} {\bibfnamefont {B.}~\bibnamefont
  {Kundys}},\ }\href {https://doi.org/10.1063/1.4905505} {\bibfield  {journal}
  {\bibinfo  {journal} {Applied Physics Reviews}\ }\textbf {\bibinfo {volume}
  {2}},\ \bibinfo {pages} {011301} (\bibinfo {year} {2015})}\BibitemShut
  {NoStop}%
\bibitem [{\citenamefont {Matzen}\ \emph {et~al.}(2019)\citenamefont {Matzen},
  \citenamefont {Guillemot}, \citenamefont {Maroutian}, \citenamefont {Patel},
  \citenamefont {Wen}, \citenamefont {DiChiara}, \citenamefont {Agnus},
  \citenamefont {Shpyrko}, \citenamefont {Fullerton}, \citenamefont
  {Ravelosona}, \citenamefont {Lecoeur},\ and\ \citenamefont
  {Kukreja}}]{Matzen2019}%
  \BibitemOpen
  \bibfield  {author} {\bibinfo {author} {\bibfnamefont {S.}~\bibnamefont
  {Matzen}}, \bibinfo {author} {\bibfnamefont {L.}~\bibnamefont {Guillemot}},
  \bibinfo {author} {\bibfnamefont {T.}~\bibnamefont {Maroutian}}, \bibinfo
  {author} {\bibfnamefont {S.~K.~K.}\ \bibnamefont {Patel}}, \bibinfo {author}
  {\bibfnamefont {H.}~\bibnamefont {Wen}}, \bibinfo {author} {\bibfnamefont
  {A.~D.}\ \bibnamefont {DiChiara}}, \bibinfo {author} {\bibfnamefont
  {G.}~\bibnamefont {Agnus}}, \bibinfo {author} {\bibfnamefont {O.~G.}\
  \bibnamefont {Shpyrko}}, \bibinfo {author} {\bibfnamefont {E.~E.}\
  \bibnamefont {Fullerton}}, \bibinfo {author} {\bibfnamefont {D.}~\bibnamefont
  {Ravelosona}}, \bibinfo {author} {\bibfnamefont {P.}~\bibnamefont
  {Lecoeur}},\ and\ \bibinfo {author} {\bibfnamefont {R.}~\bibnamefont
  {Kukreja}},\ }\href {https://doi.org/https://doi.org/10.1002/aelm.201800709}
  {\bibfield  {journal} {\bibinfo  {journal} {Advanced Electronic Materials}\
  }\textbf {\bibinfo {volume} {5}},\ \bibinfo {pages} {1800709} (\bibinfo
  {year} {2019})}\BibitemShut {NoStop}%
\bibitem [{\citenamefont {Guillemeney}\ \emph {et~al.}(2022)\citenamefont
  {Guillemeney}, \citenamefont {Lermusiaux}, \citenamefont {Landaburu},
  \citenamefont {Wagnon},\ and\ \citenamefont
  {Ab{\'e}cassis}}]{Guillemeney2022}%
  \BibitemOpen
  \bibfield  {author} {\bibinfo {author} {\bibfnamefont {L.}~\bibnamefont
  {Guillemeney}}, \bibinfo {author} {\bibfnamefont {L.}~\bibnamefont
  {Lermusiaux}}, \bibinfo {author} {\bibfnamefont {G.}~\bibnamefont
  {Landaburu}}, \bibinfo {author} {\bibfnamefont {B.}~\bibnamefont {Wagnon}},\
  and\ \bibinfo {author} {\bibfnamefont {B.}~\bibnamefont {Ab{\'e}cassis}},\
  }\href {https://doi.org/10.1038/s42004-021-00621-z} {\bibfield  {journal}
  {\bibinfo  {journal} {Communications Chemistry}\ }\textbf {\bibinfo {volume}
  {5}},\ \bibinfo {pages} {7} (\bibinfo {year} {2022})}\BibitemShut {NoStop}%
\bibitem [{\citenamefont {Prinz}\ \emph {et~al.}(2000)\citenamefont {Prinz},
  \citenamefont {Seleznev}, \citenamefont {Gutakovsky}, \citenamefont
  {Chehovskiy}, \citenamefont {Preobrazhenskii}, \citenamefont {Putyato},\ and\
  \citenamefont {Gavrilova}}]{Prinz2000}%
  \BibitemOpen
  \bibfield  {author} {\bibinfo {author} {\bibfnamefont {V.}~\bibnamefont
  {Prinz}}, \bibinfo {author} {\bibfnamefont {V.}~\bibnamefont {Seleznev}},
  \bibinfo {author} {\bibfnamefont {A.}~\bibnamefont {Gutakovsky}}, \bibinfo
  {author} {\bibfnamefont {A.}~\bibnamefont {Chehovskiy}}, \bibinfo {author}
  {\bibfnamefont {V.}~\bibnamefont {Preobrazhenskii}}, \bibinfo {author}
  {\bibfnamefont {M.}~\bibnamefont {Putyato}},\ and\ \bibinfo {author}
  {\bibfnamefont {T.}~\bibnamefont {Gavrilova}},\ }\href
  {https://doi.org/https://doi.org/10.1016/S1386-9477(99)00249-0} {\bibfield
  {journal} {\bibinfo  {journal} {Physica E: Low-dimensional Systems and
  Nanostructures}\ }\textbf {\bibinfo {volume} {6}},\ \bibinfo {pages} {828}
  (\bibinfo {year} {2000})}\BibitemShut {NoStop}%
\bibitem [{\citenamefont {Schmidt}\ and\ \citenamefont
  {Eberl}(2001)}]{schmidt2001}%
  \BibitemOpen
  \bibfield  {author} {\bibinfo {author} {\bibfnamefont {O.~G.}\ \bibnamefont
  {Schmidt}}\ and\ \bibinfo {author} {\bibfnamefont {K.}~\bibnamefont
  {Eberl}},\ }\href {https://doi.org/10.1038/35065525} {\bibfield  {journal}
  {\bibinfo  {journal} {Nature}\ }\textbf {\bibinfo {volume} {410}},\ \bibinfo
  {pages} {168} (\bibinfo {year} {2001})}\BibitemShut {NoStop}%
\bibitem [{\citenamefont {Sahoo}\ \emph {et~al.}(2018)\citenamefont {Sahoo},
  \citenamefont {Mondal}, \citenamefont {Williams}, \citenamefont {May},
  \citenamefont {Ladak},\ and\ \citenamefont {Barman}}]{Sahoo2018}%
  \BibitemOpen
  \bibfield  {author} {\bibinfo {author} {\bibfnamefont {S.}~\bibnamefont
  {Sahoo}}, \bibinfo {author} {\bibfnamefont {S.}~\bibnamefont {Mondal}},
  \bibinfo {author} {\bibfnamefont {G.}~\bibnamefont {Williams}}, \bibinfo
  {author} {\bibfnamefont {A.}~\bibnamefont {May}}, \bibinfo {author}
  {\bibfnamefont {S.}~\bibnamefont {Ladak}},\ and\ \bibinfo {author}
  {\bibfnamefont {A.}~\bibnamefont {Barman}},\ }\href
  {https://doi.org/10.1039/C7NR07843A} {\bibfield  {journal} {\bibinfo
  {journal} {Nanoscale}\ }\textbf {\bibinfo {volume} {10}},\ \bibinfo {pages}
  {9981} (\bibinfo {year} {2018})}\BibitemShut {NoStop}%
\bibitem [{\citenamefont {Gibbs}\ \emph {et~al.}(2014)\citenamefont {Gibbs},
  \citenamefont {Mark}, \citenamefont {Lee}, \citenamefont {Eslami},
  \citenamefont {Schamel},\ and\ \citenamefont {Fischer}}]{Gibbs2014}%
  \BibitemOpen
  \bibfield  {author} {\bibinfo {author} {\bibfnamefont {J.~G.}\ \bibnamefont
  {Gibbs}}, \bibinfo {author} {\bibfnamefont {A.~G.}\ \bibnamefont {Mark}},
  \bibinfo {author} {\bibfnamefont {T.-C.}\ \bibnamefont {Lee}}, \bibinfo
  {author} {\bibfnamefont {S.}~\bibnamefont {Eslami}}, \bibinfo {author}
  {\bibfnamefont {D.}~\bibnamefont {Schamel}},\ and\ \bibinfo {author}
  {\bibfnamefont {P.}~\bibnamefont {Fischer}},\ }\href
  {https://doi.org/10.1039/C4NR00403E} {\bibfield  {journal} {\bibinfo
  {journal} {Nanoscale}\ }\textbf {\bibinfo {volume} {6}},\ \bibinfo {pages}
  {9457} (\bibinfo {year} {2014})}\BibitemShut {NoStop}%
\bibitem [{\citenamefont {Dobrovolskiy}\ \emph {et~al.}(2021)\citenamefont
  {Dobrovolskiy}, \citenamefont {Vovk}, \citenamefont {Bondarenko},
  \citenamefont {Bunyaev}, \citenamefont {Lamb-Camarena}, \citenamefont
  {Zenbaa}, \citenamefont {Sachser}, \citenamefont {Barth}, \citenamefont
  {Guslienko}, \citenamefont {Chumak}, \citenamefont {Huth},\ and\
  \citenamefont {Kakazei}}]{Dobrovolskiy2021}%
  \BibitemOpen
  \bibfield  {author} {\bibinfo {author} {\bibfnamefont {O.~V.}\ \bibnamefont
  {Dobrovolskiy}}, \bibinfo {author} {\bibfnamefont {N.~R.}\ \bibnamefont
  {Vovk}}, \bibinfo {author} {\bibfnamefont {A.~V.}\ \bibnamefont
  {Bondarenko}}, \bibinfo {author} {\bibfnamefont {S.~A.}\ \bibnamefont
  {Bunyaev}}, \bibinfo {author} {\bibfnamefont {S.}~\bibnamefont
  {Lamb-Camarena}}, \bibinfo {author} {\bibfnamefont {N.}~\bibnamefont
  {Zenbaa}}, \bibinfo {author} {\bibfnamefont {R.}~\bibnamefont {Sachser}},
  \bibinfo {author} {\bibfnamefont {S.}~\bibnamefont {Barth}}, \bibinfo
  {author} {\bibfnamefont {K.~Y.}\ \bibnamefont {Guslienko}}, \bibinfo {author}
  {\bibfnamefont {A.~V.}\ \bibnamefont {Chumak}}, \bibinfo {author}
  {\bibfnamefont {M.}~\bibnamefont {Huth}},\ and\ \bibinfo {author}
  {\bibfnamefont {G.~N.}\ \bibnamefont {Kakazei}},\ }\href
  {https://doi.org/10.1063/5.0044325} {\bibfield  {journal} {\bibinfo
  {journal} {Applied Physics Letters}\ }\textbf {\bibinfo {volume} {118}},\
  \bibinfo {pages} {132405} (\bibinfo {year} {2021})}\BibitemShut {NoStop}%
\bibitem [{\citenamefont {Sanz-Hern{\'a}ndez}\ \emph
  {et~al.}(2020)\citenamefont {Sanz-Hern{\'a}ndez}, \citenamefont
  {Hierro-Rodriguez}, \citenamefont {Donnelly}, \citenamefont {Pablo-Navarro},
  \citenamefont {Sorrentino}, \citenamefont {Pereiro}, \citenamefont
  {Mag{\'e}n}, \citenamefont {McVitie}, \citenamefont {de~Teresa},
  \citenamefont {Ferrer}, \citenamefont {Fischer},\ and\ \citenamefont
  {Fern{\'a}ndez-Pacheco}}]{Sanz-Hernandez2020}%
  \BibitemOpen
  \bibfield  {author} {\bibinfo {author} {\bibfnamefont {D.}~\bibnamefont
  {Sanz-Hern{\'a}ndez}}, \bibinfo {author} {\bibfnamefont {A.}~\bibnamefont
  {Hierro-Rodriguez}}, \bibinfo {author} {\bibfnamefont {C.}~\bibnamefont
  {Donnelly}}, \bibinfo {author} {\bibfnamefont {J.}~\bibnamefont
  {Pablo-Navarro}}, \bibinfo {author} {\bibfnamefont {A.}~\bibnamefont
  {Sorrentino}}, \bibinfo {author} {\bibfnamefont {E.}~\bibnamefont {Pereiro}},
  \bibinfo {author} {\bibfnamefont {C.}~\bibnamefont {Mag{\'e}n}}, \bibinfo
  {author} {\bibfnamefont {S.}~\bibnamefont {McVitie}}, \bibinfo {author}
  {\bibfnamefont {J.~M.}\ \bibnamefont {de~Teresa}}, \bibinfo {author}
  {\bibfnamefont {S.}~\bibnamefont {Ferrer}}, \bibinfo {author} {\bibfnamefont
  {P.}~\bibnamefont {Fischer}},\ and\ \bibinfo {author} {\bibfnamefont
  {A.}~\bibnamefont {Fern{\'a}ndez-Pacheco}},\ }\href
  {https://doi.org/10.1021/acsnano.0c00720} {\bibfield  {journal} {\bibinfo
  {journal} {ACS Nano}\ }\textbf {\bibinfo {volume} {14}},\ \bibinfo {pages}
  {8084} (\bibinfo {year} {2020})}\BibitemShut {NoStop}%
\bibitem [{\citenamefont {Skoric}\ \emph {et~al.}(2020)\citenamefont {Skoric},
  \citenamefont {Sanz-Hern{\'a}ndez}, \citenamefont {Meng}, \citenamefont
  {Donnelly}, \citenamefont {Merino-Aceituno},\ and\ \citenamefont
  {Fern{\'a}ndez-Pacheco}}]{Skoric2020}%
  \BibitemOpen
  \bibfield  {author} {\bibinfo {author} {\bibfnamefont {L.}~\bibnamefont
  {Skoric}}, \bibinfo {author} {\bibfnamefont {D.}~\bibnamefont
  {Sanz-Hern{\'a}ndez}}, \bibinfo {author} {\bibfnamefont {F.}~\bibnamefont
  {Meng}}, \bibinfo {author} {\bibfnamefont {C.}~\bibnamefont {Donnelly}},
  \bibinfo {author} {\bibfnamefont {S.}~\bibnamefont {Merino-Aceituno}},\ and\
  \bibinfo {author} {\bibfnamefont {A.}~\bibnamefont {Fern{\'a}ndez-Pacheco}},\
  }\href {https://doi.org/10.1021/acs.nanolett.9b03565} {\bibfield  {journal}
  {\bibinfo  {journal} {Nano Letters}\ }\textbf {\bibinfo {volume} {20}},\
  \bibinfo {pages} {184} (\bibinfo {year} {2020})}\BibitemShut {NoStop}%
\bibitem [{\citenamefont {Castillo-Sep\'ulveda}\ \emph
  {et~al.}(2017)\citenamefont {Castillo-Sep\'ulveda}, \citenamefont {Escobar},
  \citenamefont {Altbir}, \citenamefont {Krizanac},\ and\ \citenamefont
  {Vedmedenko}}]{CastilloSepulveda2017}%
  \BibitemOpen
  \bibfield  {author} {\bibinfo {author} {\bibfnamefont {S.}~\bibnamefont
  {Castillo-Sep\'ulveda}}, \bibinfo {author} {\bibfnamefont {R.~A.}\
  \bibnamefont {Escobar}}, \bibinfo {author} {\bibfnamefont {D.}~\bibnamefont
  {Altbir}}, \bibinfo {author} {\bibfnamefont {M.}~\bibnamefont {Krizanac}},\
  and\ \bibinfo {author} {\bibfnamefont {E.~Y.}\ \bibnamefont {Vedmedenko}},\
  }\href {https://doi.org/10.1103/PhysRevB.96.024426} {\bibfield  {journal}
  {\bibinfo  {journal} {Phys. Rev. B}\ }\textbf {\bibinfo {volume} {96}},\
  \bibinfo {pages} {024426} (\bibinfo {year} {2017})}\BibitemShut {NoStop}%
\bibitem [{\citenamefont {Pylypovskyi}\ \emph {et~al.}(2020)\citenamefont
  {Pylypovskyi}, \citenamefont {Kononenko}, \citenamefont {Yershov},
  \citenamefont {R{\"o}{\ss}ler}, \citenamefont {Tomilo}, \citenamefont
  {Fassbender}, \citenamefont {van~den Brink}, \citenamefont {Makarov},\ and\
  \citenamefont {Sheka}}]{Pylypovskyi2020}%
  \BibitemOpen
  \bibfield  {author} {\bibinfo {author} {\bibfnamefont {O.~V.}\ \bibnamefont
  {Pylypovskyi}}, \bibinfo {author} {\bibfnamefont {D.~Y.}\ \bibnamefont
  {Kononenko}}, \bibinfo {author} {\bibfnamefont {K.~V.}\ \bibnamefont
  {Yershov}}, \bibinfo {author} {\bibfnamefont {U.~K.}\ \bibnamefont
  {R{\"o}{\ss}ler}}, \bibinfo {author} {\bibfnamefont {A.~V.}\ \bibnamefont
  {Tomilo}}, \bibinfo {author} {\bibfnamefont {J.}~\bibnamefont {Fassbender}},
  \bibinfo {author} {\bibfnamefont {J.}~\bibnamefont {van~den Brink}}, \bibinfo
  {author} {\bibfnamefont {D.}~\bibnamefont {Makarov}},\ and\ \bibinfo {author}
  {\bibfnamefont {D.~D.}\ \bibnamefont {Sheka}},\ }\href
  {https://doi.org/10.1021/acs.nanolett.0c03246} {\bibfield  {journal}
  {\bibinfo  {journal} {Nano Letters}\ }\textbf {\bibinfo {volume} {20}},\
  \bibinfo {pages} {8157} (\bibinfo {year} {2020})}\BibitemShut {NoStop}%
\bibitem [{\citenamefont {Yershov}(2022)}]{Yershov2022}%
  \BibitemOpen
  \bibfield  {author} {\bibinfo {author} {\bibfnamefont {K.~V.}\ \bibnamefont
  {Yershov}},\ }\href {https://doi.org/10.1103/PhysRevB.105.064407} {\bibfield
  {journal} {\bibinfo  {journal} {Phys. Rev. B}\ }\textbf {\bibinfo {volume}
  {105}},\ \bibinfo {pages} {064407} (\bibinfo {year} {2022})}\BibitemShut
  {NoStop}%
\bibitem [{\citenamefont {Borysenko}\ \emph {et~al.}(2022)\citenamefont
  {Borysenko}, \citenamefont {Sheka}, \citenamefont {Fassbender}, \citenamefont
  {van~den Brink}, \citenamefont {Makarov},\ and\ \citenamefont
  {Pylypovskyi}}]{Borysenko2022}%
  \BibitemOpen
  \bibfield  {author} {\bibinfo {author} {\bibfnamefont {Y.~A.}\ \bibnamefont
  {Borysenko}}, \bibinfo {author} {\bibfnamefont {D.~D.}\ \bibnamefont
  {Sheka}}, \bibinfo {author} {\bibfnamefont {J.}~\bibnamefont {Fassbender}},
  \bibinfo {author} {\bibfnamefont {J.}~\bibnamefont {van~den Brink}}, \bibinfo
  {author} {\bibfnamefont {D.}~\bibnamefont {Makarov}},\ and\ \bibinfo {author}
  {\bibfnamefont {O.~V.}\ \bibnamefont {Pylypovskyi}},\ }\href
  {https://doi.org/10.1103/PhysRevB.106.174426} {\bibfield  {journal} {\bibinfo
   {journal} {Phys. Rev. B}\ }\textbf {\bibinfo {volume} {106}},\ \bibinfo
  {pages} {174426} (\bibinfo {year} {2022})}\BibitemShut {NoStop}%
\bibitem [{\citenamefont {Makarov}\ \emph {et~al.}(2022)\citenamefont
  {Makarov}, \citenamefont {Volkov}, \citenamefont {Kákay}, \citenamefont
  {Pylypovskyi}, \citenamefont {Budinská},\ and\ \citenamefont
  {Dobrovolskiy}}]{Makarov2022}%
  \BibitemOpen
  \bibfield  {author} {\bibinfo {author} {\bibfnamefont {D.}~\bibnamefont
  {Makarov}}, \bibinfo {author} {\bibfnamefont {O.~M.}\ \bibnamefont {Volkov}},
  \bibinfo {author} {\bibfnamefont {A.}~\bibnamefont {Kákay}}, \bibinfo
  {author} {\bibfnamefont {O.~V.}\ \bibnamefont {Pylypovskyi}}, \bibinfo
  {author} {\bibfnamefont {B.}~\bibnamefont {Budinská}},\ and\ \bibinfo
  {author} {\bibfnamefont {O.~V.}\ \bibnamefont {Dobrovolskiy}},\ }\href
  {https://doi.org/https://doi.org/10.1002/adma.202101758} {\bibfield
  {journal} {\bibinfo  {journal} {Advanced Materials}\ }\textbf {\bibinfo
  {volume} {34}},\ \bibinfo {pages} {2101758} (\bibinfo {year}
  {2022})}\BibitemShut {NoStop}%
\bibitem [{\citenamefont {Bergeret}\ \emph {et~al.}(2001)\citenamefont
  {Bergeret}, \citenamefont {Volkov},\ and\ \citenamefont
  {Efetov}}]{Bergeret2001}%
  \BibitemOpen
  \bibfield  {author} {\bibinfo {author} {\bibfnamefont {F.~S.}\ \bibnamefont
  {Bergeret}}, \bibinfo {author} {\bibfnamefont {A.~F.}\ \bibnamefont
  {Volkov}},\ and\ \bibinfo {author} {\bibfnamefont {K.~B.}\ \bibnamefont
  {Efetov}},\ }\href {https://doi.org/10.1103/PhysRevLett.86.4096} {\bibfield
  {journal} {\bibinfo  {journal} {Phys. Rev. Lett.}\ }\textbf {\bibinfo
  {volume} {86}},\ \bibinfo {pages} {4096} (\bibinfo {year}
  {2001})}\BibitemShut {NoStop}%
\bibitem [{\citenamefont {Khaire}\ \emph {et~al.}(2010)\citenamefont {Khaire},
  \citenamefont {Khasawneh}, \citenamefont {Pratt},\ and\ \citenamefont
  {Birge}}]{Khaire2010}%
  \BibitemOpen
  \bibfield  {author} {\bibinfo {author} {\bibfnamefont {T.~S.}\ \bibnamefont
  {Khaire}}, \bibinfo {author} {\bibfnamefont {M.~A.}\ \bibnamefont
  {Khasawneh}}, \bibinfo {author} {\bibfnamefont {W.~P.}\ \bibnamefont
  {Pratt}},\ and\ \bibinfo {author} {\bibfnamefont {N.~O.}\ \bibnamefont
  {Birge}},\ }\href {https://doi.org/10.1103/PhysRevLett.104.137002} {\bibfield
   {journal} {\bibinfo  {journal} {Phys. Rev. Lett.}\ }\textbf {\bibinfo
  {volume} {104}},\ \bibinfo {pages} {137002} (\bibinfo {year}
  {2010})}\BibitemShut {NoStop}%
\bibitem [{\citenamefont {Robinson}\ \emph {et~al.}(2010)\citenamefont
  {Robinson}, \citenamefont {Witt},\ and\ \citenamefont
  {Blamire}}]{Robinson2010}%
  \BibitemOpen
  \bibfield  {author} {\bibinfo {author} {\bibfnamefont {J.~W.~A.}\
  \bibnamefont {Robinson}}, \bibinfo {author} {\bibfnamefont {J.~D.~S.}\
  \bibnamefont {Witt}},\ and\ \bibinfo {author} {\bibfnamefont {M.~G.}\
  \bibnamefont {Blamire}},\ }\href {https://doi.org/10.1126/science.1189246}
  {\bibfield  {journal} {\bibinfo  {journal} {Science}\ }\textbf {\bibinfo
  {volume} {329}},\ \bibinfo {pages} {59} (\bibinfo {year} {2010})}\BibitemShut
  {NoStop}%
\bibitem [{\citenamefont {Bergeret}\ and\ \citenamefont
  {Tokatly}(2013)}]{Bergeret2013}%
  \BibitemOpen
  \bibfield  {author} {\bibinfo {author} {\bibfnamefont {F.~S.}\ \bibnamefont
  {Bergeret}}\ and\ \bibinfo {author} {\bibfnamefont {I.~V.}\ \bibnamefont
  {Tokatly}},\ }\href {https://doi.org/10.1103/PhysRevLett.110.117003}
  {\bibfield  {journal} {\bibinfo  {journal} {Phys. Rev. Lett.}\ }\textbf
  {\bibinfo {volume} {110}},\ \bibinfo {pages} {117003} (\bibinfo {year}
  {2013})}\BibitemShut {NoStop}%
\bibitem [{\citenamefont {Bergeret}\ and\ \citenamefont
  {Tokatly}(2014)}]{Bergeret2014}%
  \BibitemOpen
  \bibfield  {author} {\bibinfo {author} {\bibfnamefont {F.~S.}\ \bibnamefont
  {Bergeret}}\ and\ \bibinfo {author} {\bibfnamefont {I.~V.}\ \bibnamefont
  {Tokatly}},\ }\href {https://doi.org/10.1103/PhysRevB.89.134517} {\bibfield
  {journal} {\bibinfo  {journal} {Physical Review B}\ }\textbf {\bibinfo
  {volume} {89}},\ \bibinfo {pages} {134517} (\bibinfo {year}
  {2014})}\BibitemShut {NoStop}%
\bibitem [{\citenamefont {Bulaevskii}\ \emph {et~al.}(2017)\citenamefont
  {Bulaevskii}, \citenamefont {Eneias},\ and\ \citenamefont
  {Ferraz}}]{Bulaevskii2017}%
  \BibitemOpen
  \bibfield  {author} {\bibinfo {author} {\bibfnamefont {L.}~\bibnamefont
  {Bulaevskii}}, \bibinfo {author} {\bibfnamefont {R.}~\bibnamefont {Eneias}},\
  and\ \bibinfo {author} {\bibfnamefont {A.}~\bibnamefont {Ferraz}},\ }\href
  {https://doi.org/10.1103/PhysRevB.95.104513} {\bibfield  {journal} {\bibinfo
  {journal} {Phys. Rev. B}\ }\textbf {\bibinfo {volume} {95}},\ \bibinfo
  {pages} {104513} (\bibinfo {year} {2017})}\BibitemShut {NoStop}%
\bibitem [{\citenamefont {Jakobsen}\ \emph {et~al.}(2020)\citenamefont
  {Jakobsen}, \citenamefont {Naess}, \citenamefont {Dutta}, \citenamefont
  {Brataas},\ and\ \citenamefont {Qaiumzadeh}}]{Jakobsen2020}%
  \BibitemOpen
  \bibfield  {author} {\bibinfo {author} {\bibfnamefont {M.~F.}\ \bibnamefont
  {Jakobsen}}, \bibinfo {author} {\bibfnamefont {K.~B.}\ \bibnamefont {Naess}},
  \bibinfo {author} {\bibfnamefont {P.}~\bibnamefont {Dutta}}, \bibinfo
  {author} {\bibfnamefont {A.}~\bibnamefont {Brataas}},\ and\ \bibinfo {author}
  {\bibfnamefont {A.}~\bibnamefont {Qaiumzadeh}},\ }\href
  {https://doi.org/10.1103/PhysRevB.102.140504} {\bibfield  {journal} {\bibinfo
   {journal} {Phys. Rev. B}\ }\textbf {\bibinfo {volume} {102}},\ \bibinfo
  {pages} {140504(R)} (\bibinfo {year} {2020})}\BibitemShut {NoStop}%
\bibitem [{\citenamefont {Jakobsen}\ \emph {et~al.}(2021)\citenamefont
  {Jakobsen}, \citenamefont {Brataas},\ and\ \citenamefont
  {Qaiumzadeh}}]{Jakobsen2021}%
  \BibitemOpen
  \bibfield  {author} {\bibinfo {author} {\bibfnamefont {M.~F.}\ \bibnamefont
  {Jakobsen}}, \bibinfo {author} {\bibfnamefont {A.}~\bibnamefont {Brataas}},\
  and\ \bibinfo {author} {\bibfnamefont {A.}~\bibnamefont {Qaiumzadeh}},\
  }\href {https://doi.org/10.1103/PhysRevLett.127.017701} {\bibfield  {journal}
  {\bibinfo  {journal} {Phys. Rev. Lett.}\ }\textbf {\bibinfo {volume} {127}},\
  \bibinfo {pages} {017701} (\bibinfo {year} {2021})}\BibitemShut {NoStop}%
\bibitem [{\citenamefont {Johnsen}\ \emph {et~al.}(2021)\citenamefont
  {Johnsen}, \citenamefont {Jacobsen},\ and\ \citenamefont
  {Linder}}]{Johnsen2021}%
  \BibitemOpen
  \bibfield  {author} {\bibinfo {author} {\bibfnamefont {L.~G.}\ \bibnamefont
  {Johnsen}}, \bibinfo {author} {\bibfnamefont {S.~H.}\ \bibnamefont
  {Jacobsen}},\ and\ \bibinfo {author} {\bibfnamefont {J.}~\bibnamefont
  {Linder}},\ }\href {https://doi.org/10.1103/PhysRevB.103.L060505} {\bibfield
  {journal} {\bibinfo  {journal} {Phys. Rev. B}\ }\textbf {\bibinfo {volume}
  {103}},\ \bibinfo {pages} {L060505} (\bibinfo {year} {2021})}\BibitemShut
  {NoStop}%
\bibitem [{\citenamefont {Hauser}\ \emph {et~al.}(1966)\citenamefont {Hauser},
  \citenamefont {Theuerer},\ and\ \citenamefont {Werthamer}}]{Hauser1966}%
  \BibitemOpen
  \bibfield  {author} {\bibinfo {author} {\bibfnamefont {J.~J.}\ \bibnamefont
  {Hauser}}, \bibinfo {author} {\bibfnamefont {H.~C.}\ \bibnamefont
  {Theuerer}},\ and\ \bibinfo {author} {\bibfnamefont {N.~R.}\ \bibnamefont
  {Werthamer}},\ }\href {https://doi.org/10.1103/PhysRev.142.118} {\bibfield
  {journal} {\bibinfo  {journal} {Phys. Rev.}\ }\textbf {\bibinfo {volume}
  {142}},\ \bibinfo {pages} {118} (\bibinfo {year} {1966})}\BibitemShut
  {NoStop}%
\bibitem [{\citenamefont {Zhen}\ \emph {et~al.}(2019)\citenamefont {Zhen},
  \citenamefont {Zhang}, \citenamefont {Zhang},\ and\ \citenamefont
  {Dong}}]{Zhen2019}%
  \BibitemOpen
  \bibfield  {author} {\bibinfo {author} {\bibfnamefont {S.}~\bibnamefont
  {Zhen}}, \bibinfo {author} {\bibfnamefont {H.}~\bibnamefont {Zhang}},
  \bibinfo {author} {\bibfnamefont {Q.}~\bibnamefont {Zhang}},\ and\ \bibinfo
  {author} {\bibfnamefont {Z.}~\bibnamefont {Dong}},\ }\href
  {https://doi.org/10.1007/s10948-018-4960-9} {\bibfield  {journal} {\bibinfo
  {journal} {Journal of Superconductivity and Novel Magnetism}\ }\textbf
  {\bibinfo {volume} {32}},\ \bibinfo {pages} {1945} (\bibinfo {year}
  {2019})}\BibitemShut {NoStop}%
\bibitem [{\citenamefont {Hübener}\ \emph {et~al.}(2002)\citenamefont
  {Hübener}, \citenamefont {Tikhonov}, \citenamefont {Garifullin},
  \citenamefont {Westerholt},\ and\ \citenamefont {Zabel}}]{Hubener2002}%
  \BibitemOpen
  \bibfield  {author} {\bibinfo {author} {\bibfnamefont {M.}~\bibnamefont
  {Hübener}}, \bibinfo {author} {\bibfnamefont {D.}~\bibnamefont {Tikhonov}},
  \bibinfo {author} {\bibfnamefont {I.~A.}\ \bibnamefont {Garifullin}},
  \bibinfo {author} {\bibfnamefont {K.}~\bibnamefont {Westerholt}},\ and\
  \bibinfo {author} {\bibfnamefont {H.}~\bibnamefont {Zabel}},\ }\href
  {https://doi.org/10.1088/0953-8984/14/37/305} {\bibfield  {journal} {\bibinfo
   {journal} {Journal of Physics: Condensed Matter}\ }\textbf {\bibinfo
  {volume} {14}},\ \bibinfo {pages} {8687} (\bibinfo {year}
  {2002})}\BibitemShut {NoStop}%
\bibitem [{\citenamefont {Bobkov}\ \emph {et~al.}(2022)\citenamefont {Bobkov},
  \citenamefont {Bobkova}, \citenamefont {Bobkov},\ and\ \citenamefont
  {Kamra}}]{Bobkov2022}%
  \BibitemOpen
  \bibfield  {author} {\bibinfo {author} {\bibfnamefont {G.~A.}\ \bibnamefont
  {Bobkov}}, \bibinfo {author} {\bibfnamefont {I.~V.}\ \bibnamefont {Bobkova}},
  \bibinfo {author} {\bibfnamefont {A.~M.}\ \bibnamefont {Bobkov}},\ and\
  \bibinfo {author} {\bibfnamefont {A.}~\bibnamefont {Kamra}},\ }\href
  {https://doi.org/10.1103/PhysRevB.106.144512} {\bibfield  {journal} {\bibinfo
   {journal} {Phys. Rev. B}\ }\textbf {\bibinfo {volume} {106}},\ \bibinfo
  {pages} {144512} (\bibinfo {year} {2022})}\BibitemShut {NoStop}%
\bibitem [{\citenamefont {Fyhn}\ \emph
  {et~al.}(2023{\natexlab{b}})\citenamefont {Fyhn}, \citenamefont {Brataas},
  \citenamefont {Qaiumzadeh},\ and\ \citenamefont {Linder}}]{Fyhn2023}%
  \BibitemOpen
  \bibfield  {author} {\bibinfo {author} {\bibfnamefont {E.~H.}\ \bibnamefont
  {Fyhn}}, \bibinfo {author} {\bibfnamefont {A.}~\bibnamefont {Brataas}},
  \bibinfo {author} {\bibfnamefont {A.}~\bibnamefont {Qaiumzadeh}},\ and\
  \bibinfo {author} {\bibfnamefont {J.}~\bibnamefont {Linder}},\ }\href
  {https://doi.org/10.1103/PhysRevB.107.174503} {\bibfield  {journal} {\bibinfo
   {journal} {Phys. Rev. B}\ }\textbf {\bibinfo {volume} {107}},\ \bibinfo
  {pages} {174503} (\bibinfo {year} {2023}{\natexlab{b}})}\BibitemShut
  {NoStop}%
\bibitem [{\citenamefont {Bobkov}\ \emph {et~al.}(2023)\citenamefont {Bobkov},
  \citenamefont {Bobkova},\ and\ \citenamefont {Bobkov}}]{Bobkov2023}%
  \BibitemOpen
  \bibfield  {author} {\bibinfo {author} {\bibfnamefont {G.~A.}\ \bibnamefont
  {Bobkov}}, \bibinfo {author} {\bibfnamefont {I.~V.}\ \bibnamefont
  {Bobkova}},\ and\ \bibinfo {author} {\bibfnamefont {A.~M.}\ \bibnamefont
  {Bobkov}},\ }\href {https://doi.org/10.1103/PhysRevB.108.054510} {\bibfield
  {journal} {\bibinfo  {journal} {Physical Review B}\ }\textbf {\bibinfo
  {volume} {108}},\ \bibinfo {pages} {054510} (\bibinfo {year}
  {2023})}\BibitemShut {NoStop}%
\bibitem [{\citenamefont {Ortix}(2015)}]{ortix2015quantum}%
  \BibitemOpen
  \bibfield  {author} {\bibinfo {author} {\bibfnamefont {C.}~\bibnamefont
  {Ortix}},\ }\href {https://doi.org/10.1103/PhysRevB.91.245412} {\bibfield
  {journal} {\bibinfo  {journal} {Phys. Rev. B}\ }\textbf {\bibinfo {volume}
  {91}},\ \bibinfo {pages} {245412} (\bibinfo {year} {2015})}\BibitemShut
  {NoStop}%
\bibitem [{\citenamefont {Kelly}(2021)}]{Kelly2020}%
  \BibitemOpen
  \bibfield  {author} {\bibinfo {author} {\bibfnamefont {P.~A.}\ \bibnamefont
  {Kelly}},\ }in\ \href
  {https://pkel015.connect.amazon.auckland.ac.nz/SolidMechanicsBooks/Part_III/index.html}
  {\emph {\bibinfo {booktitle} {Mechanics Lecture Notes Part III: Foundations
  of Continuum Mechanics}}}\ (\bibinfo  {publisher} {University of Auckland},\
  \bibinfo {year} {2021})\ Chap.~\bibinfo {chapter} {1}\BibitemShut {NoStop}%
\bibitem [{\citenamefont {Machon}\ \emph {et~al.}(2013)\citenamefont {Machon},
  \citenamefont {Eschrig},\ and\ \citenamefont {Belzig}}]{Machon2013}%
  \BibitemOpen
  \bibfield  {author} {\bibinfo {author} {\bibfnamefont {P.}~\bibnamefont
  {Machon}}, \bibinfo {author} {\bibfnamefont {M.}~\bibnamefont {Eschrig}},\
  and\ \bibinfo {author} {\bibfnamefont {W.}~\bibnamefont {Belzig}},\ }\href
  {https://doi.org/10.1103/PhysRevLett.110.047002} {\bibfield  {journal}
  {\bibinfo  {journal} {Phys. Rev. Lett.}\ }\textbf {\bibinfo {volume} {110}},\
  \bibinfo {pages} {047002} (\bibinfo {year} {2013})}\BibitemShut {NoStop}%
\bibitem [{\citenamefont {Kuprianov}\ and\ \citenamefont
  {Lukichev}(1988)}]{KuprianovLukichev1988}%
  \BibitemOpen
  \bibfield  {author} {\bibinfo {author} {\bibfnamefont {M.~Y.}\ \bibnamefont
  {Kuprianov}}\ and\ \bibinfo {author} {\bibfnamefont {V.}~\bibnamefont
  {Lukichev}},\ }\href@noop {} {\bibfield  {journal} {\bibinfo  {journal} {Zh.
  Eksp. Teor. Fiz}\ }\textbf {\bibinfo {volume} {94}},\ \bibinfo {pages} {149}
  (\bibinfo {year} {1988})}\BibitemShut {NoStop}%
\bibitem [{\citenamefont {Jacobsen}\ \emph {et~al.}(2015)\citenamefont
  {Jacobsen}, \citenamefont {Ouassou},\ and\ \citenamefont
  {Linder}}]{Jacobsen2015b}%
  \BibitemOpen
  \bibfield  {author} {\bibinfo {author} {\bibfnamefont {S.~H.}\ \bibnamefont
  {Jacobsen}}, \bibinfo {author} {\bibfnamefont {J.~A.}\ \bibnamefont
  {Ouassou}},\ and\ \bibinfo {author} {\bibfnamefont {J.}~\bibnamefont
  {Linder}},\ }\href {https://doi.org/10.1103/PhysRevB.92.024510} {\bibfield
  {journal} {\bibinfo  {journal} {Physical Review B}\ }\textbf {\bibinfo
  {volume} {92}},\ \bibinfo {pages} {024510} (\bibinfo {year}
  {2015})}\BibitemShut {NoStop}%
\bibitem [{\citenamefont {Schopohl}\ and\ \citenamefont
  {Maki}(1995)}]{Schopohl1995}%
  \BibitemOpen
  \bibfield  {author} {\bibinfo {author} {\bibfnamefont {N.}~\bibnamefont
  {Schopohl}}\ and\ \bibinfo {author} {\bibfnamefont {K.}~\bibnamefont
  {Maki}},\ }\href {https://doi.org/10.1103/PhysRevB.52.490} {\bibfield
  {journal} {\bibinfo  {journal} {Phys. Rev. B}\ }\textbf {\bibinfo {volume}
  {52}},\ \bibinfo {pages} {490} (\bibinfo {year} {1995})}\BibitemShut
  {NoStop}%
\bibitem [{\citenamefont {Tanaka}\ and\ \citenamefont
  {Golubov}(2007)}]{Tanaka2007}%
  \BibitemOpen
  \bibfield  {author} {\bibinfo {author} {\bibfnamefont {Y.}~\bibnamefont
  {Tanaka}}\ and\ \bibinfo {author} {\bibfnamefont {A.~A.}\ \bibnamefont
  {Golubov}},\ }\href {https://doi.org/10.1103/PhysRevLett.98.037003}
  {\bibfield  {journal} {\bibinfo  {journal} {Phys. Rev. Lett.}\ }\textbf
  {\bibinfo {volume} {98}},\ \bibinfo {pages} {037003} (\bibinfo {year}
  {2007})}\BibitemShut {NoStop}%
\bibitem [{\citenamefont {Kontos}\ \emph {et~al.}(2001)\citenamefont {Kontos},
  \citenamefont {Aprili}, \citenamefont {Lesueur},\ and\ \citenamefont
  {Grison}}]{Kontos2001}%
  \BibitemOpen
  \bibfield  {author} {\bibinfo {author} {\bibfnamefont {T.}~\bibnamefont
  {Kontos}}, \bibinfo {author} {\bibfnamefont {M.}~\bibnamefont {Aprili}},
  \bibinfo {author} {\bibfnamefont {J.}~\bibnamefont {Lesueur}},\ and\ \bibinfo
  {author} {\bibfnamefont {X.}~\bibnamefont {Grison}},\ }\href
  {https://doi.org/10.1103/PhysRevLett.86.304} {\bibfield  {journal} {\bibinfo
  {journal} {Phys. Rev. Lett.}\ }\textbf {\bibinfo {volume} {86}},\ \bibinfo
  {pages} {304} (\bibinfo {year} {2001})}\BibitemShut {NoStop}%
\bibitem [{Zhang \emph{et~al.}(2016)}]{Zhang2016}
\BibitemOpen
\bibfield {author} {X. Zhang, B. Li,\ and\ J. Zhang,\ }\href
  {https://doi.org/10.1021/acs.inorgchem.5b02785}
{\bibfield  {journal} {\bibinfo
  {journal} {Inorg. Chem.}\ }\textbf {\bibinfo {volume} {55}},\ \bibinfo {issue} {7},\ \bibinfo
  {pages} {3378-3383} (\bibinfo {year} {2016})}
\BibitemShut{NoStop}%
\bibitem [{Samanta \emph{et~al.}(2014)}]{Samanta2014}
\BibitemOpen
\bibfield {author} {P. K. Samanta\ and\ S. K. Pati,\ }\href
  {https://doi.org/10.1002/chem.201302628}
{\bibfield  {journal} {\bibinfo
  {journal} {Chem. Eur. J.}\ }\textbf {\bibinfo {volume} {20}},\ \bibinfo
  {pages} {1760-1764} (\bibinfo {year} {2014})}
\BibitemShut{NoStop}%
\end{thebibliography}

%

\end{document}